\documentclass[preprint,showpacs,titlepage,aps,prd,tightenlines,amsmath,byrevtex,nofootinbib]{revtex4}
\pdfoutput=1
\usepackage{natbib}
\usepackage{epsfig}
\usepackage{graphicx}
\usepackage{dcolumn}
\usepackage{amsmath}
\usepackage{amssymb}
\usepackage{enumerate}
\usepackage{subfigure}    
\usepackage{multirow}
\usepackage{color} 

\newcommand{\thab}{\theta_{12}}
\newcommand{\thac}{\theta_{13}}
\newcommand{\thbc}{\theta_{23}}

\newcommand{\dms}{\Delta m_{21}^{2}}
\newcommand{\dma}{\Delta m_{31}^{2}}

\newcommand{\nue}{\nu_{e}}
\newcommand{\numu}{\nu_{\mu}}

\begin{document}

\vskip-6pt \hfill {DCPT/11/136, IPPP/11/68, IFT-UAM/CSIC-11-72, IFIC/11-52} \\

\title{A comparative study of long-baseline superbeams within LAGUNA for large $\theta_{13}$}
\author{Pilar Coloma}
\email{pcoloma@vt.edu}
\affiliation{Instituto de F\'isica Te\'orica UAM/CSIC, Cantoblanco 28049, Madrid, Spain \& Center for Neutrino Physics, Department of Physics, Virginia Tech, Blacksburg, 24061 VA, USA}
\author{Tracey Li}
\email{tracey.li@ific.uv.es}
\affiliation{Instituto de F\'isica Corpuscular, Edificio Institutos de Investigaci\'on, Paterna, Valencia E-46071, Spain}
\author{Silvia Pascoli}
\email{silvia.pascoli@durham.ac.uk}
\affiliation{IPPP, Department of Physics, Durham University, Durham DH1 3LE, United Kingdom}


\begin{abstract}

The Daya Bay and RENO experiments have recently observed a non-zero $\theta_{13}$ at more than $5\sigma$ CL. This has important consequences for future neutrino oscillation experiments. We analyze these within the LAGUNA design study which considers seven possible locations for a European neutrino observatory for proton decay, neutrino, and astroparticle physics. The megaton-scale detector would be an ideal target for a CERN-based neutrino beam with baselines ranging from 130 km to 2300 km. We perform a detailed study to assess the physics reach of the three detector options - a 440 kton water \v{C}erenkov, a 100 kton liquid argon and a 50 kton liquid scintillator detector - at each of the possible locations, taking into account the recent measurement of $\theta_{13}$. We study the impact of the beam properties and detector performances on the sensitivity to CP-violation and the mass hierarchy. We find that a liquid argon or water \v{C}erenkov detector can make a $3\sigma$ discovery of CP violation for $60\%-70\%$ of the parameter space for any of the baselines under consideration, although the results for the liquid argon detector placed at 130 km are slightly worse and only $40\%-50\%$ is achieved in this case. The performance of the liquid scintillator detector is affected by its level of neutral-current background at all baselines. A $3\sigma$ determination of the mass hierarchy is possible for all values of $\delta$, for the values of $\theta_{13}$ favoured at $3\sigma$ by Daya Bay and RENO, for almost all setups with $L\gtrsim 650$ km.

\end{abstract}


\maketitle

\section{Introduction}

Neutrino oscillations have been established over the past decade by numerous experiments, providing experimental evidence for the existence of new physics beyond the Standard Model (SM), needed to explain non-zero neutrino masses and mixing in the leptonic sector. In the standard three-family framework, neutrino oscillations are parameterised by three mixing angles ($\thab,\,\thbc,\,\thac$), two mass-squared differences ($\dms,\,\dma$, defined as $\Delta m^2_{ij}\equiv m_i^2-m_j^2 $) and a CP-violating phase\footnote{If neutrinos are Majorana particles, there are two additional CP-violating phases but these do not enter oscillation experiments.}, $\delta$. The best-fit values for the oscillation parameters, based on the global fit from Ref.~\cite{Schwetz:2011qt}, are $\sin^{2}\theta_{12}=0.312$, $\sin^{2}\theta_{23}=0.51\,(0.52)$, $\dms=7.59\times10^{-5}\mbox{eV}^{2}$ and $\dma=2.45\times10^{-3}\mbox{ eV}^{2}$ ($-2.34\times10^{-3}\mbox{ eV}^{2}$) for a normal (inverted) hierarchy. A very recently global analysis finds very similar results~\cite{Fogli:2012ua}. The neutrino mass hierarchy and the value of the CP phase $\delta$ are as yet unknown. The only constraint on $\delta$ comes from a tentative bound from the Super-Kamiokande experiment~\cite{Obayashi:2010zz} and a first indication of $\delta \sim \pi$ at 1$\sigma$ from the global analysis in Ref.~\cite{Fogli:2012ua}. 

It is the third mixing angle, $\thac$, that has attracted much attention in the past year. First came the results obtained from the T2K experiment~\cite{Abe:2011sj} in June 2011 studying the $\numu\to\nue$ appearance channel, followed shortly by the results from the MINOS experiment~\cite{Adamson:2011qu} for the same channel. This was followed by the results from the Double-Chooz collaboration for the $\bar\nu_e$ disappearance channel~\cite{Abe:2011fz} in December 2011. All these results seemed to point towards a non-zero value of $\thac$ at a rather low statistical significance, although combined global fits give evidence above the $3\sigma$ CL~\cite{Fogli:2011qn,Schwetz:2011zk}). These hints have now been confirmed by the Daya Bay and the RENO collaborations, who have recently reported the discovery of a non-zero $\theta_{13}$~\cite{An:2012eh, collaboration:2012nd}. Both best fit values lie in the same range, around $\sin^22\theta_{13}\sim 0.09-0.10$. The combined global fit from Ref.~\cite{Fogli:2012ua} gives $0.0214<\sin^2\theta_{13}<0.0279$ at $1\sigma$ CL for normal hierarchy, with its best fit at $\sin^2\theta_{13}=0.0245$. Very similar results are obtained for an inverted hierarchy.

This impressive result has crucial consequences from both theoretical and phenomenological perspectives. On the one hand, precise measurements of the mixing parameters in the leptonic sector (and similarly in the quark sector) are crucial in order to search for the underlying theory behind the flavour structure of the SM. The indication that $\thac\neq 0$ is a vital piece of information from the point of view of model-building and the search for flavour symmetries. On the other hand, a non-zero $\theta_{13}$ opens up the possibility of observing CP violation in the leptonic sector and of measuring the ordering of the mass eigenstates. Future experiments are currently being designed in order to observe oscillations mainly in the $\nue\to\numu$ or $\numu\to\nue$ channels (together with their CP-conjugate channels) which depend simultaneously on $\thac,\,\delta$ and the mass hierarchy. The value of $\thac$ has always dictated how future experiments must be optimised in order to measure the unknown oscillation parameters - the fact that $\thac$ is known to be large now means that whilst it is still important to maximise statistics and minimise backgrounds, more work must be directed at minimising the systematic errors which will ultimately limit the performance of the experiment. 

Given this relatively large value of $\theta_{13}$, it may be possible to observe CP violation and measure the mass hierarchy with a superbeam~\cite{Richter:2000pu}, a $\beta$-beam~\cite{Zucchelli:2002sa} (or a combination of the two~\cite{Campagne:2006yx}) or a Low Energy Neutrino Factory~\cite{Geer:2007kn,Bross:2007ts,FernandezMartinez:2010zza,Agarwalla:2010hk} aimed at a large-volume detector with good background rejection capabilities and energy resolution. 

The recently completed LAGUNA (Large Apparatus for Grand Unification and Neutrino Astrophysics)~\cite{Autiero:2007zj,laguna,Rubbia:2010fm,Rubbia:2010zz} design study was a European effort to evaluate the feasibility of seven selected sites, listed in Tab.~\ref{tab:sites}, to host a giant, deep-underground particle detector. Three possible options are currently envisioned for the detector technology: 100 kton of Liquid Argon (LAGUNA LAr or GLACIER~\cite{Rubbia:2009md}), 50 kton of Liquid Scintillator (LAGUNA LSc or LENA~\cite{MarrodanUndagoitia:2008zz,Wurm:2011zn}) or 440 kton of Water \v{C}erenkov (LAGUNA WC or MEMPHYS~\cite{deBellefon:2006vq}). 
The detector will be a multi-purpose facility able to detect proton decay and neutrinos from both natural and man-made sources, making it an obvious candidate for  the far detector in a next-generation neutrino beam experiment. In this context, the LAGUNA-LBNO (Long-Baseline Neutrino Oscillation)~\cite{Rubbia:2010zz} study will evaluate the feasibility of the production of a neutrino superbeam at CERN. In terms of the prospective LAGUNA sites, a crucial factor is then the distance between each of these sites and CERN, as this dictates the required peak energy of the beam in order to match the first oscillation maximum and maximise the signal rate. 

Several studies of $\beta$-beams aimed from CERN to one of the LAGUNA sites have already been performed~\cite{Mezzetto:2003ub,BurguetCastell:2003vv,Mezzetto:2004gs,BurguetCastell:2005pa, Campagne:2006yx,Donini:2006tt,Donini:2006dx,Donini:2007qt,Meloni:2008it,FernandezMartinez:2009hb,Orme:2010cc,Coloma:2010wa}. However, technical difficulties related to ion production, collection and acceleration make the $\beta$-beam option very challenging. Therefore, in this work we will focus on the physics reach of superbeams instead, since the technology needed to produce them is relatively well known and will probably be at hand in the near future. 

In this paper we evaluate and compare the performances of all the possible combinations of the seven prospective baselines and three detectors from a phenomenological perspective, regarding the sensitivity of the different setups to the neutrino oscillation parameters $\delta$, the mass hierarchy, and non-maximal $\theta_{23}$ ($\theta_{12}\neq45^{\circ}$) in light of the recent measurement of $\theta_{13}$. We also study in detail the impact of the beam and detector properties on the performance of the setups, thereby determining the most important factors and demonstrating the necessity of obtaining accurate information on both the beam and detector sides in order to perform realistic simulations and objective comparisons. 

The paper is structured as follows. In Sec.~\ref{sec:sim} we describe our simulation method, explaining our assumptions about the beam and the detectors, as well as the details of the numerical analysis. We then present in Sec.~\ref{sec:opt} our results for the impact of several experimental factors on the performance of the experiment: systematic errors, backgrounds, the running times in neutrino/anti-neutrino modes, and $\tau$ detection. In Sec.~\ref{sec:results} we directly compare the performances of all the prospective baselines and detectors, in terms of the discovery potentials for CP violation, the mass hierarchy and non-maximal $\theta_{23}$. In Sec.~\ref{sec:exposure} we consider how the performance varies with the total exposure (running time $\times$ detector mass $\times$ power), comparing the results for the shortest and longest baselines in terms of the CPV discovery potential and the $\theta_{13}-\delta$ precision as a function of exposure. Finally, in Sec.~\ref{sec:conclusions} we summarise and draw our conclusions.

\begin{center}
\begin{table}
\renewcommand{\arraystretch}{1.8}
\begin{tabular}{ c | c | c }
\hline
\textbf{Location} & \textbf{Distance from CERN [km]} & \textbf{1st osc max [GeV]}\\ \hline \hline
Fr\'ejus (France)     & 130   & 0.26 \\
Canfranc (Spain)      & 630   & 1.27 \\
Umbria (Italy)        & 665   & 1.34 \\
Sierozsowice (Poland) & 950   & 1.92 \\
Boulby (UK)           & 1050  & 2.12 \\
Slanic (Romania)      & 1570  & 3.18 \\
Pyh\"asalmi (Finland) & 2300  & 4.65 \\ \hline
\end{tabular}\\
\caption{The seven potential sites under consideration in the LAGUNA design study. The energy of the first oscillation maximum is calculated in the absence of matter effects. From Ref.~\cite{Rubbia:2010fm}.}
\label{tab:sites}
\end{table}
\end{center}


\section{Simulation details}
\label{sec:sim}

We perform a numerical simulation of the experimental setups using the publicly available software package General Long-Baseline Experiment Simulator (GLoBES) \cite{globes1,globes2}. For the matter density, we use the Preliminary Reference Earth Model (PREM) profile, calculated by GLoBES from Refs.~\cite{matter1,matter2}, together with an uncertainty of $2\%$\footnote{We have checked that a $5\%$ matter uncertainty does not significantly affect the results.}~\cite{density}. The neutrino interaction cross-sections have been taken from Refs.~\cite{Messier:1999kj,Paschos:2001np}. The detector simulation is explained in detail in the following subsection. We have included constant systematic uncertainties over the signal and the background rates, which have been included as normalization errors. Therefore these are correlated between different energy bins for a given channel, but are uncorrelated between different channels. Unless otherwise stated, systematic errors have been taken at the $5\%$ level for both the signal and background rates for all setups, regardless of the detector technology. 

The true values for the solar and atmospheric parameters have been set as: $\theta_{12}=34.2^\circ$, $\theta_{23}=45^\circ$, $\Delta m_{21}^2 = 7.59\times 10^{-5}$ eV$^{-5}$ and $\Delta m_{31}^2 = 2.45\times 10^{-3}$ eV$^{-3}$, in agreement with the best-fit values from Ref.~\cite{Schwetz:2011qt}. Our results will generally be presented in the region $0.01<\sin^{2}2\theta_{13}<0.1$, in view of the Daya Bay and RENO results. Marginalization has been performed assuming $4\%$ and $10\%$ gaussian priors centered around these values, for the solar and atmospheric parameters respectively. A normal mass hierarchy has been assumed for all the results shown in this paper; the results for the inverted hierarchy are similar with the exchange $\delta\rightarrow -\delta$. Unless otherwise stated, our results will be presented in terms of the $3\sigma$ discovery potential, i.e. the ability to exclude a given hypothesis at $3\sigma$ confidence level (1 d.o.f.), for CP-conservation (corresponding to $\delta = 0, \pi$), the wrong mass hierarchy, and maximal $\theta_{23}$ ($\theta_{23} = 45^\circ$). 

For the CPV discovery plots, we separate the $\delta$-parameter space into two regions ($\delta\leq0$ and $\delta\geq0$) rather than including the entire region in one plot, so that the details of the differences between the setups can be better appreciated.

\subsection{The beam}\label{sec:beam}
 
Superbeam fluxes, optimised for each baseline according to the first oscillation peak, have been provided by A.~Longhin~\cite{longhin,Longhin:2010zz,Longhin:2011hn}. In the lowest energy configuration of the beam (for $L = 130$ km), the fluxes have been obtained assuming $5.6\times10^{22}$ protons on target (PoT) per year, with an energy of 4.5 GeV~\cite{Longhin:2011hn}. We assume 2 year of $\nu$ running and 8 years of $\bar{\nu}$ running in this case, as in Ref.~\cite{Campagne:2006yx}. Higher energy fluxes~\cite{Longhin:2010zz} (in the multi-GeV regime, for baselines with $L > 130$ km) correspond to the CERN high-power PS2 configuration~\cite{Rubbia:2010fm}: $3\times10^{21}$ PoT per year, with an energy of 50 GeV. In this case we assume 5 years of $\nu$ running and 5 years of $\bar{\nu}$, as will be discussed in Section~\ref{sec:times}. It is important to notice that the two beam configurations correspond to different beam powers: 4 (2.7) MW for the CERN-Frej\'us beam and 2.4 (1.6) MW for the other baselines, assuming $1.0~(1.5) \times 10^7$ useful seconds per year. Therefore, the comparison between the physics reach of the different setups should be considered carefully, see Sec.~\ref{sec:results}. In the work presented here we are mostly interested in studying these agressive scenarios, with multi-MW beams. In Sec.~\ref{sec:exposure} we show how the results for CPV would vary if the total exposure is decreased (if, for instance, the beam power or the mass of the detector are reduced).

\begin{figure}[htp]
\centering
\subfigure[~$\nu$ beam spectrum for 2300 km.]{\label{fig:nu_beam}
          \includegraphics[scale=0.6]{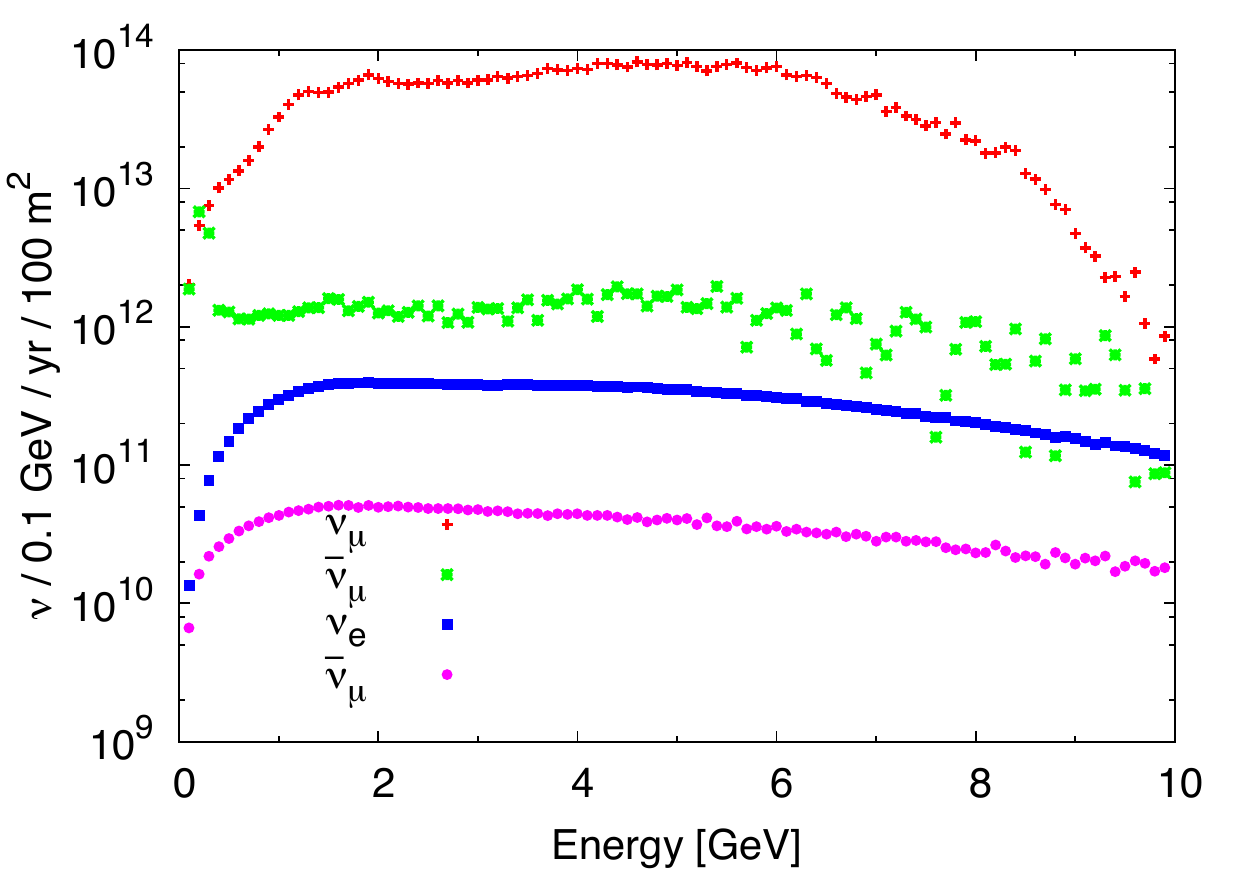}}
\subfigure[~$\nu_{\mu}$ content of different beams.]{\label{fig:flux_baselines}
          \includegraphics[scale=0.6]{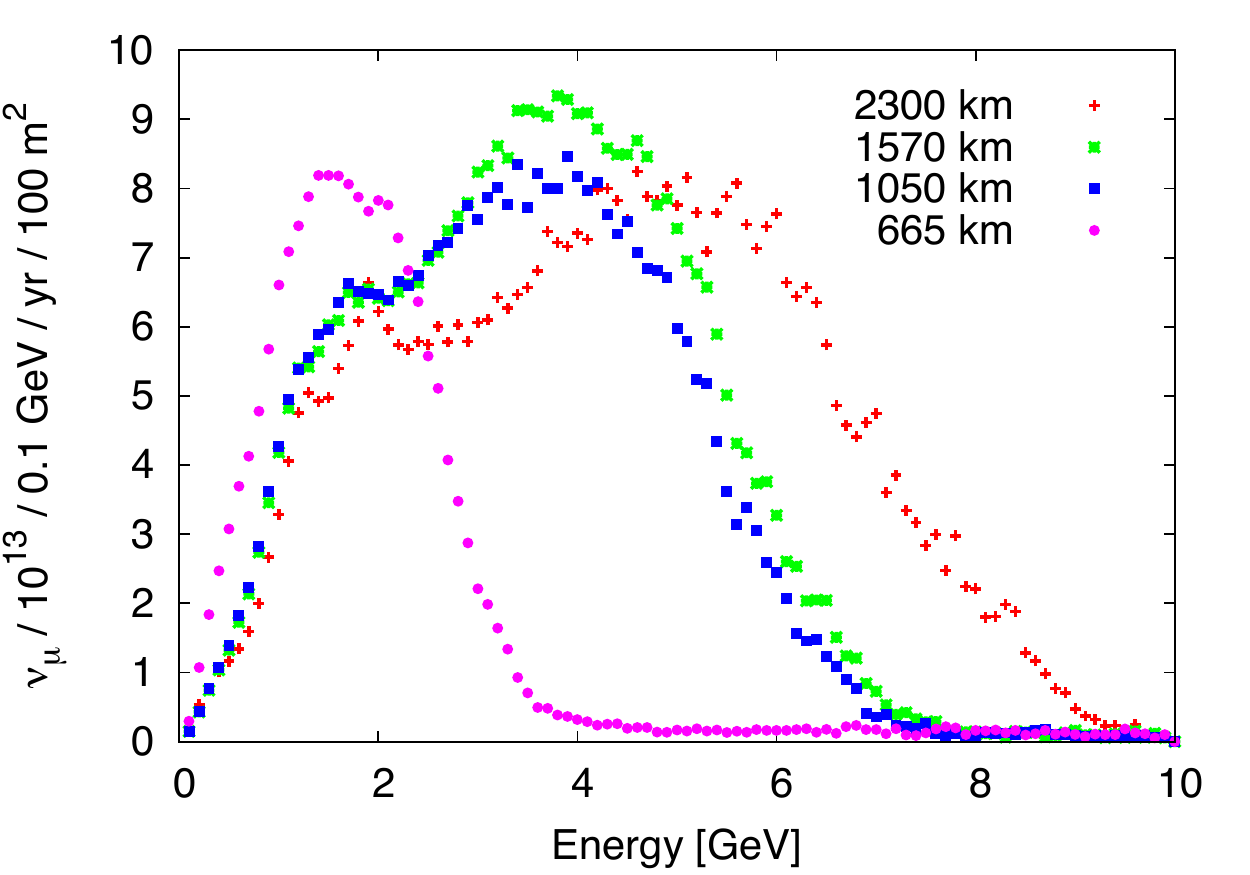}}\\
\caption{a) Content of the $\nu$ beam optimised for 2300 km and b) comparison of the $\nu_{\mu}$ content of the $\nu$ beams optimised for different baselines. Fluxes have been provided by A.~Longhin~\cite{longhin}.}
\label{fig:beam}
\end{figure}

As an example, the spectra of the $\nu$ beam, optimised for the 2300 km baseline, is shown in Fig.~\ref{fig:nu_beam}. In addition to the main $\nu_{\mu}$ ($\bar{\nu}_{\mu}$) content, the beam also contains $\sim4\%$ contamination from $\bar\nu_{\mu}$ ($\nu_{\mu}$), $\sim1\%$ contamination from $\nu_{e}$ ($\bar\nu_{e}$) and $\sim 0.1\%$ from $\bar{\nu}_{e}$ ($\nu_{e}$) which act as irreducible backgrounds. The effect of these backgrounds on the performance of the experiment will be discussed further in Sec.~\ref{sec:intr}. The composition of the $\bar{\nu}$ beam is similar. In Fig.~\ref{fig:flux_baselines} we show the $\nu_{\mu}$ content of the beams optimised for 2300 km, 1570 km, 1050 km and 665 km. The beam for 950 km is very similar to that for 1050 km, and that for 650 km is very similar to that for 665 km. From this figure it can be seen that the higher energy beams generally have a broader peak than the lower energy beams. The spectrum of the 130 km beam can be found in Ref.~\cite{Longhin:2011hn}.
 
\subsection{The detectors}\label{sec:detectors}

Three detectors are currently under consideration within the LAGUNA-LBNO study:

\begin{description}

\item[LAGUNA LAr or GLACIER] (Giant Liquid Argon Charge Imaging ExpeRiment~\cite{Rubbia:2009md}) is a 100 kton liquid argon detector using the Liquid Argon Time Projection Chamber (LAr TPC) technology. It has excellent particle identification and energy resolution, a low energy threshold (of a few MeV for electrons and a few tens of MeV for protons), good suppression of NC background events (which produce pion decays at the detector), and good signal efficiencies~\cite{LBNEreport,Barger:2007jq}. 

\item[LAGUNA LSc or LENA] (Low Energy Neutrino Astronomy~\cite{MarrodanUndagoitia:2008zz,Wurm:2011zn}) is a 50 kton Liquid Scintillator (LSc) detector optimised for the detection of low-energy neutrinos. It is based on the same technology developed for the BOREXINO detector~\cite{Arpesella:2007xf}. It has good energy resolution, a very low energy threshold, and high light yield. Its use as a target for a superbeam experiment in the 100~MeV--few~GeV energy range has been recently proposed~\cite{randolph,randolph2,Wurm:2011zn} and is currently under intense study.

\item[LAGUNA WC or MEMPHYS] (MEgaton Mass PHYSics~\cite{deBellefon:2006vq}) is a 440 kton Water \v{C}erenkov (WC) detector\footnote{In LAGUNA-LBNO, a revised version of the WC detector is being considered with an increased size to 500 kton. We expect the results to improve slightly with respect to the ones presented here, due to the consequent increase in the number of events.}. Three detector designs for Mton WC detectors are currently being carried out: Hyper-Kamiokande in Japan~\cite{Nakamura:2003hk}, UNO in the USA~\cite{Jung:1999jq} and MEMPHYS in Europe~\cite{deBellefon:2006vq}. The WC detector technology is the cheapest and most suitable for instrumenting very massive detectors. However, WC detectors are not optimal for high-energy neutrino interactions because of the difficulties of using the multi-ring events that appear at higher energies, and of distinguishing electrons from neutral pions that may be produced from neutral-current interactions.

\end{description}

The detector details are summarised in Tab.~\ref{tab:detectors}. Unless stated otherwise, these are the values that we use in all our simulations. 

\begin{table}[hbtp]
\renewcommand{\arraystretch}{1.8}
\begin{center}
\begin{tabular}{|c|c|c|c|c|c|c|}
\hline 
  Detector & M (kton) & $\epsilon $  & NC bckgr. & $\sigma(E)$ & $E_\nu$ (GeV) \\
\hline
\hline
 LAr  ($L=130$)   & 100   & 90 & 0.5\% & Matr.~\cite{LArmigrationmatrices} & $\left[0.1,1\right]$   \\

 \multirow{2}{*}{LAr  ($L>130$)}   & \multirow{2}{*}{100}   & \multirow{2}{*}{90} & \multirow{2}{*}{0.5\%} & $0.20 E$ ($\nu_\mu$) & \multirow{2}{*}{$\left[0.1,10\right]$} \cr  
   &  &  &   & 150 MeV $(\nu_e)$  & \\ \hline
   LSc    &  50     &  $50\%$  &  $10\%$ & $0.05 E$ &  $\left[0.5,7\right]$  \\ \hline 
 WC  ($L=130$)   & \multirow{2}{*}{440}   & $\sim70\%$ & $<0.1\%$~\cite{Campagne:2006yx} & Matr.~\cite{Campagne:2006yx} & $\left[0.1,1\right]$   \cr
 WC  ($L>130$)  &  & $\sim 40\%$  & $<1\%$~\cite{Lisa}  & Matr.~\cite{Lisa}  & $\left[0.5,10\right]$ \\
\hline 
%
\hline
\end{tabular}
\end{center}
\caption{Parameters used in the simulations of each of the LAGUNA detectors. From left to right, each column corresponds to: detector technology, fiducial mass, reconstruction efficiency, percentage of NC events that are misidentified as CC events, energy resolution, and neutrino energy range. Migration matrices have been used in some cases to implement the detector response, as indicated. The efficiencies quoted for the WC detector apply to electron events, whereas for the LAr and LSc detectors these efficiencies are the same for muons and electrons. The reconstruction efficiencies for muons in the QE regime at the WC detector have been set to $97\%$.  }
\label{tab:detectors}
\end{table}

Both the LAr and the WC have been implemented differently for the $L=130$ km and the $L>130$ km baselines, due to the very different neutrino energies involved in the two cases. For the LAr detector, a constant $90\%$ efficiency is used in all cases. However, different energy resolutions have been used, depending on the baseline considered. For all baselines with $L>130$ km, the migration matrices provided by  L.~Esposito and A.~Rubbia \cite{LArmigrationmatrices} have been used for the signal. However, these matrices are optimized for high energy setups and therefore are not optimal for the lower energy setup corresponding to $L=130$ km. Therefore, in this case we have used the energy resolution functions from Ref.~\cite{Akiri:2011dv} (see Tab.~\ref{tab:detectors}).  

In order to implement the WC in the low energy regime ($L=130$ km) we have used the same efficiencies, backgrounds and migration matrices as in Ref.~\cite{Campagne:2006yx}. The details to simulate the WC exposed to higher energy beams ($L>130$ km) have been obtained from Refs.~\cite{Akiri:2011dv,brajesh}, with migration matrices for signal and background kindly provided by the LBNE collaboration~\cite{Lisa}. It should be noticed that the results presented for the WC detector placed at the intermediate baselines (namely, $L\sim 650$ km) are not optimal, as there are no migration matrices available in the literature to simulate the detector response when exposed to a superbeam in this energy range ($\sim 1.2$ GeV). We have used the same parameters as for the high energy setup in this case. This is expected to be a rather conservative approach, mainly because of two reasons: (1) the migration matrices for the high energy configurations have a low energy threshold at $0.5$ GeV; (2) the cuts on the signal needed to suppress the NC backgrounds at low energies for the LBNE setup, with $\langle E_\nu \rangle \sim 3$ GeV, may be relaxed for these setups which have much lower energies and therefore expect smaller NC backgrounds. This should be taken into account when comparing the results obtained at different baselines as it means that our results for $L\sim 650$ km are probably too pessimistic.

\subsection{Backgrounds}\label{sec:backgr}

The main background for a superbeam arises from the intrinsic contamination of the beam (see Sec.~\ref{sec:beam}). This background cannot be eliminated but various strategies can be employed to reduce its impact, e.g. by using the near detector to measure the exact composition of the flux. As can be seen from Fig.~\ref{fig:beam}, the wrong-polarity content of the beam reaches approximately the $\sim4\%$ level. Since the LAGUNA detectors are not magnetised\footnote{The possibility of a magnetised detector or a muon spectrometer to be added to the LAGUNA LAr is under consideration in LAGUNA-LBNO.}, these neutrinos constitute a background to the disappearance channel. However, the most important background is that arising from the  $\nu_{e}$ ($\bar{\nu}_{e}$) present in the beam, which is the dominant background to the primary superbeam channel, the $\nu_{\mu}\to\nu_{e}$ ($\bar{\nu}_{\mu}\to\bar{\nu}_{e}$) channel. 

The final component of the background are neutral-current (NC) events. The main source of this background are pions which are produced in NC events and are then misidentified as charged-current (CC) events, such as $\pi^\pm$'s which decay to $\mu^\pm$'s (plus missing energy). This background, which tends to pile up at low energies, affects the disappearance channel. However, if a $\pi^{0}$ is produced, it will mainly decay into two photons which, if not fully reconstructed, can be misidentified as an electron. This source of background is more problematic, since it is added to the $\nu_\mu \to \nu_e$ event rates (and its CP-conjugate). This background has been implemented in our simulations as indicated in Tab.~\ref{tab:detectors}. For the WC detector when exposed to a low energy beam, we have followed Ref.~\cite{Campagne:2006yx}, while for the higher energy configurations we have used the rejection efficiencies provided by the LBNE collaboration~\cite{Lisa}. 
In the absence of a detailed simulation for the migration of NC events to lower energies at the LAr and LSc detectors, we have assumed that a certain percentage of the unoscillated NC events that take place at the detector are mis-identified as CC events. The LAr detector has a very good efficiency for the detection of photons and it is expected that the NC background for this kind of detector will be very low. Therefore, we have assumed in this case that $0.5\%$ of the NC events produced at the detector are misidentified as CC events~\cite{CNGS, Tanimoto:2008zza}. However, the situation for the LSc detector is more problematic. The NC background rejection capabilities at this kind of detector are currently under intense study and remain unclear yet~\cite{Rubbia:2010zz}. Several cuts could help in reducing the percentage of the total sample of NC events that would constitute an actual background so that the final percentage could presumably be kept between 33\% and 11\% of the total number of NC events~\cite{Wurm}. However, these cuts would considerably reduce the efficiency to the $\nu_e$ CC signal. In the absence of an accurate estimate of both factors, we have assumed a configuration for the LSc in which a 10\% of the NC events are mis-identified as CC events, and we have assumed that the detection efficiency would be consequently reduced down to 50\%~\cite{Wurm}.  In Sec.~\ref{sec:opt} it will be shown in detail how this background affects the CPV discovery potential of the setups if a LSc detector is used. Finally, it should be noted that the inclusion of migration matrices is of crucial importance in order to evaluate the physics potential of any of the considered setups here. In the particular case of NC background events, the shape of the event sample would be very different if proper migration matrices were included in the analysis, which could affect the results.

\section{Optimisation studies}
\label{sec:opt}

In this section we study how the properties of the beam and detectors affect the sensitivity of the experiment. We have studied several variables: systematic errors, the intrinsic beam background, the NC background, the time spent running in neutrino and anti-neutrino modes, and the possibility of $\tau$ detection. Most of these studies, unless otherwise specified, have been performed in the context of a LAr detector placed at Pyh\"asalmi ($L=2300$ km).

\subsection{Systematic errors}

\begin{figure}[hbtp]
\centering
 \includegraphics[scale=0.35]{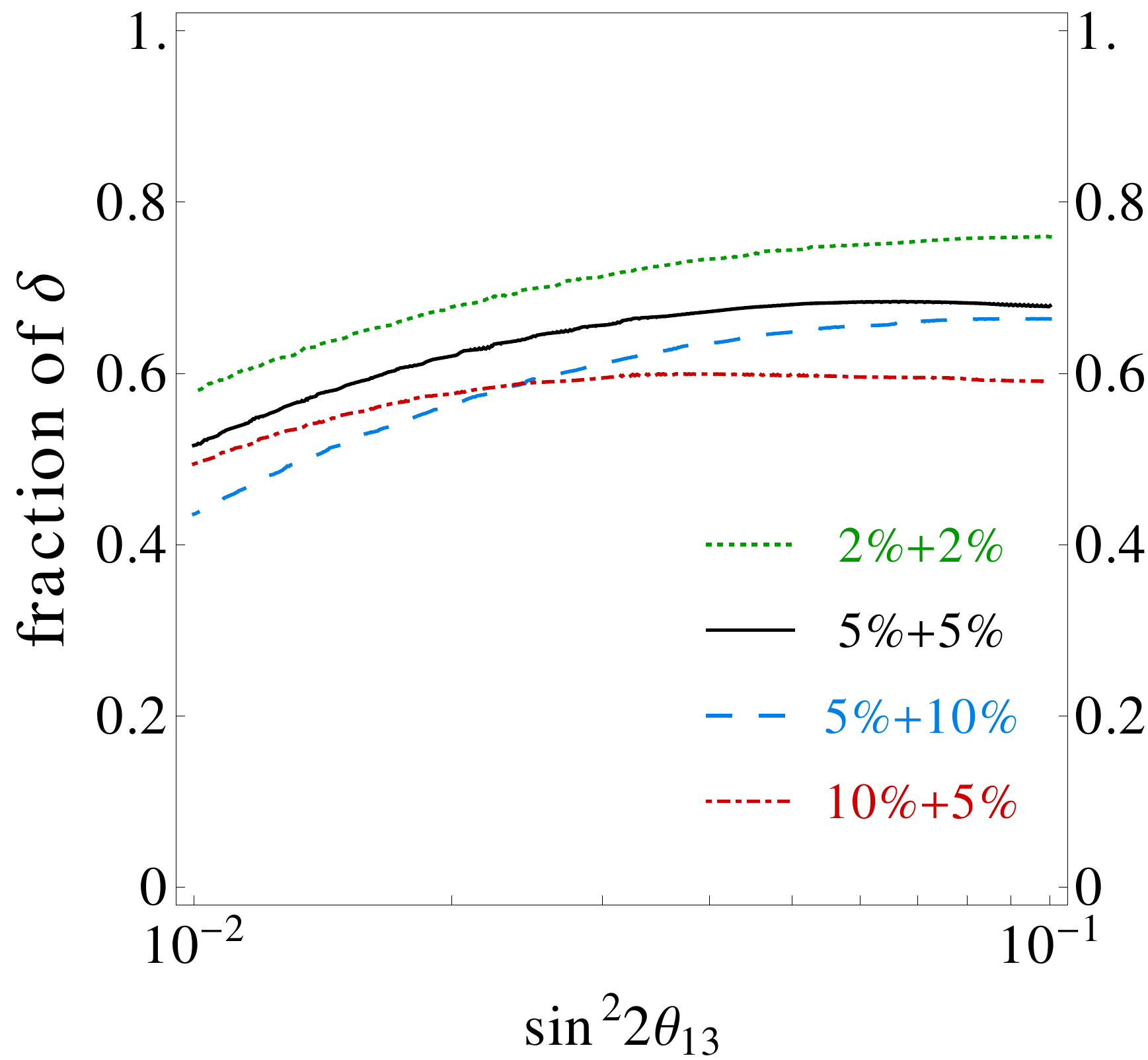}  
\caption{Effect of systematic errors on the performance of the LAr detector placed at 2300 km from the source.  The lines depict the fraction of possible values of $\delta$ for which CP violation can be established at the $3\sigma$ CL (1 d.o.f.) as a function of the true value of $\sin^22\theta_{13}$. The first value in the legend refers to the systematic error on the signal and the second to the background. The systematic errors have been taken as constant normalization errors, uncorrelated between different channels.  }
\label{fig:sys}
\end{figure}

For large $\theta_{13}$, systematic errors are the limiting factor for the performance of the experiment. Systematic errors are related to the beam and the detector and can be separated into those which affect the signal, and those which affect the predictions of the backgrounds. A dominant source comes from the estimate of the intrinsic $\nu_\mu$ and $\nu_e$ in the beam, which becomes more important at high energies due to the uncertainty on the kaon production rate. Another important source is due to the uncertainty on the detection cross-sections for the different channels. Significant improvements are expected, though, as the near detector and dedicated experiments e.g. MINERvA~\cite{Drakoulakos:2004gn} will provide crucial information. Other detector systematic errors may arise due to tracking and particle identification efficiencies and the physics modelling of the interactions. 
In the absence of a dedicated study, we let the systematic errors vary in the range from 2\% to 10\% (compared to our reference value of 5\% for all other studies in this paper) and check how they affect the CP discovery potential. In Fig.~\ref{fig:sys} we show the results for several values of the systematic uncertainties assumed for the signal and background, as indicated in the legend. The different lines depict the fraction of possible values of $\delta$ for which CP violation can be established at $3\sigma$ CL (1 d.o.f), as a function of the true value of $\sin^22\theta_{13}$. We have checked that $2\%$ systematic errors on signal and background are equivalent to no systematics at all. Results are shown for the LAr detector although a similar dependence is expected for the other detectors. 

It can be seen that CPV discovery potential is affected strongly by systematic errors. The main impact is due to the signal systematics, which reduce the CPV discovery potential of the facility in the region where $\sin^22\theta_{13} \gtrsim 2 \times 10^{-2}$. This is due to the fact that, when $\theta_{13}$ is this large, the atmospheric term (which is CP conserving, see App.~\ref{sec:probs}) dominates the oscillation probability since it is proportional to $\sin^22\theta_{13}$, while the CP violating term is subdominant since it is linear in $\sin2\theta_{13}$. Therefore, any systematic error on the signal can easily hide the CP violating contribution to the oscillation probability and deteriorate the CPV discovery potential in this regime. As a consequence,  for a $10\%$ systematic uncertainty over the signal the CP fraction would be reduced by around $\sim15\%$ around $\sin^22\theta_{13}\sim 0.09-0.10$ (the region indicated by Daya Bay and RENO) with respect to the case where there are no systematic errors. On the other hand, background systematics would become more relevant in the region where $\sin^22\theta_{13} \lesssim 2 \times 10^{-2}$ since in this regime the sensitivity to CP violation is limited by the background events. 

The effect of the systematic error for this setup is rather low compared to other superbeams in the literature (see, for instance, Ref.~\cite{Huber:2007em}). This is mainly due to two reasons: (1) our setup is limited by statistics rather than by systematics, given its very long baseline; (2) the broad flux peak, combined with a detector with very good energy resolution, makes the setup much more robust against any source of global systematic errors, since the events will be distributed into very different energy bins so the oscillation probability is better reconstructed. However, only global systematic uncertainties have been considered here. A detailed study, taking into account possible correlations between the different sources of systematic uncertainties, should be performed in order to establish their actual impact on the performance of any of the setups under consideration here.

\subsection{Intrinsic beam background}
\label{sec:intr}

As described in Section~\ref{sec:detectors}, the intrinsic beam background is one of the limiting factors of a superbeam experiment. 
From Fig.~\ref{fig:beam}, typically this background is at the sub-\% level in the energy range of interest: we see that in the $\nu$ mode (left panel), the intrinsic contamination from $\bar\nu_\mu$ is at the $\sim 4\%$ level, while the $\nu_e$ and $\bar\nu_e$ contamination is below the $1\%$ and the $0.1\%$ level, respectively. Similar levels are obtained in the $\bar\nu$ mode. We have checked how the absolute level of the background affects the results (if, for instance, the proportion of $\nu_e$ events in the beam could be reduced at the production stage), and we find that for the considered values of $\theta_{13}$ this has little effect since the signal is always large compared to this background. We have also checked that the level of intrinsic background also has virtually no effect on the precision of the measurement either, for large $\theta_{13}$.

\subsection{Neutral-current backgrounds}\label{sec:NCbck}

\begin{figure}[hbtp]
\centering
          \includegraphics[scale=0.4]{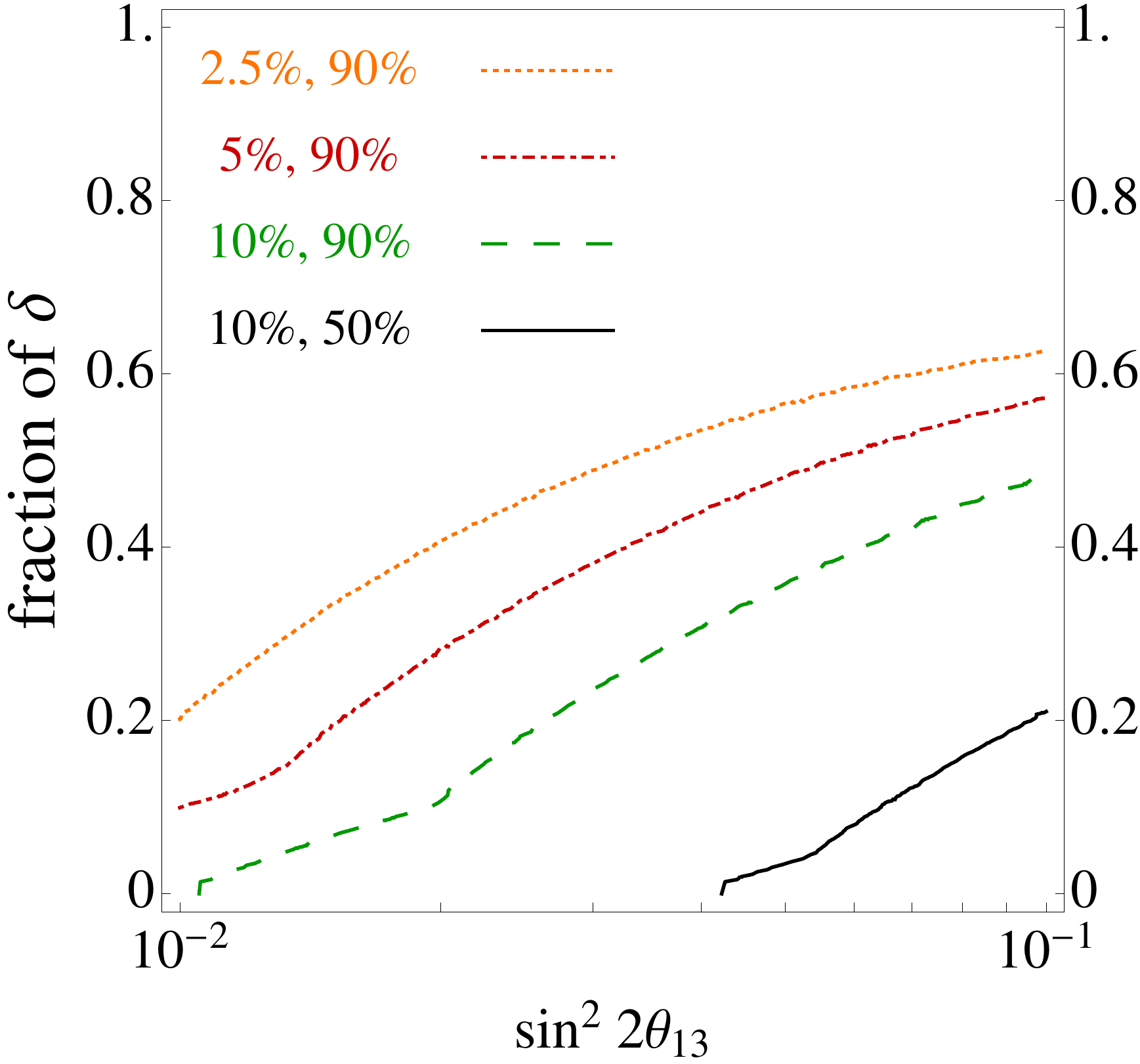}
\caption{Effect of NC background on the CPV discovery potential for the LSc detector placed at 2300 km from the source. The lines depict the fraction of possible values of delta for which CPV can be established at a $3\sigma$ CL as a function of the true value of $\sin^22\theta_{13}$ (1 d.o.f.). The first value in the legend indicates the percentage of unoscillated NC events that are misidentified as CC events, while the second value refers to the detection efficiency. For the first three cases the detection efficiency is kept at the 90\% level, while in the last case (our reference configuration for the LSc detector in the rest of this paper) it has been assumed that the cuts needed to reduce the NC background down to the 10\% level would also reduce the detection efficiency down to 50\%. }
\label{fig:NC}
\end{figure}

As mentioned in Sec.~\ref{sec:backgr}, NC backgrounds are very relevant for a superbeam experiment: $\pi^\pm$ can be misidentified as $\mu^\pm$, while two overlapping rings in the decay of a $\pi^0$ can be misidentified as an electron. The rejection capability of a LSc detector is currently unclear and in the process of being studied. In Fig.~\ref{fig:NC} we show the effect of NC backgrounds on the CPV discovery potential for this detector, for $L=2300$ km. Results are shown as a function of the true value of $\sin^22\theta_{13}$ and the fraction of possible values of $\delta$ for which CP violation can be established at the $3\sigma$ CL. We have considered that a certain percentage of the total unoscillated NC events are misidentified as CC events, as indicated in the legend. For the dotted (yellow), red (dot-dashed) and green (blue) lines it has been assumed that a reduction on the NC background could be done without affecting the CC efficiency, which has been kept at the 90\% level. However, the cuts needed to reduce this background would likely affect the detection efficiency and severely reduce it. Therefore, we also show in the same panels our reference configuration for the LSc detector (solid black lines) where only 10\% of the NC background is included in the sample, but at the price of a lower $\nu_e$ CC detection efficiency ($50$\%). These lines correspond to our reference configuration for the LSc detector (see Tab.~\ref{tab:detectors}). From the figure it can be clearly seen that the NC background levels are critical for the observation of CP violation. 

\subsection{Running times}\label{sec:times}

\begin{figure}[hbtp]
\centering
         \includegraphics[scale=0.4]{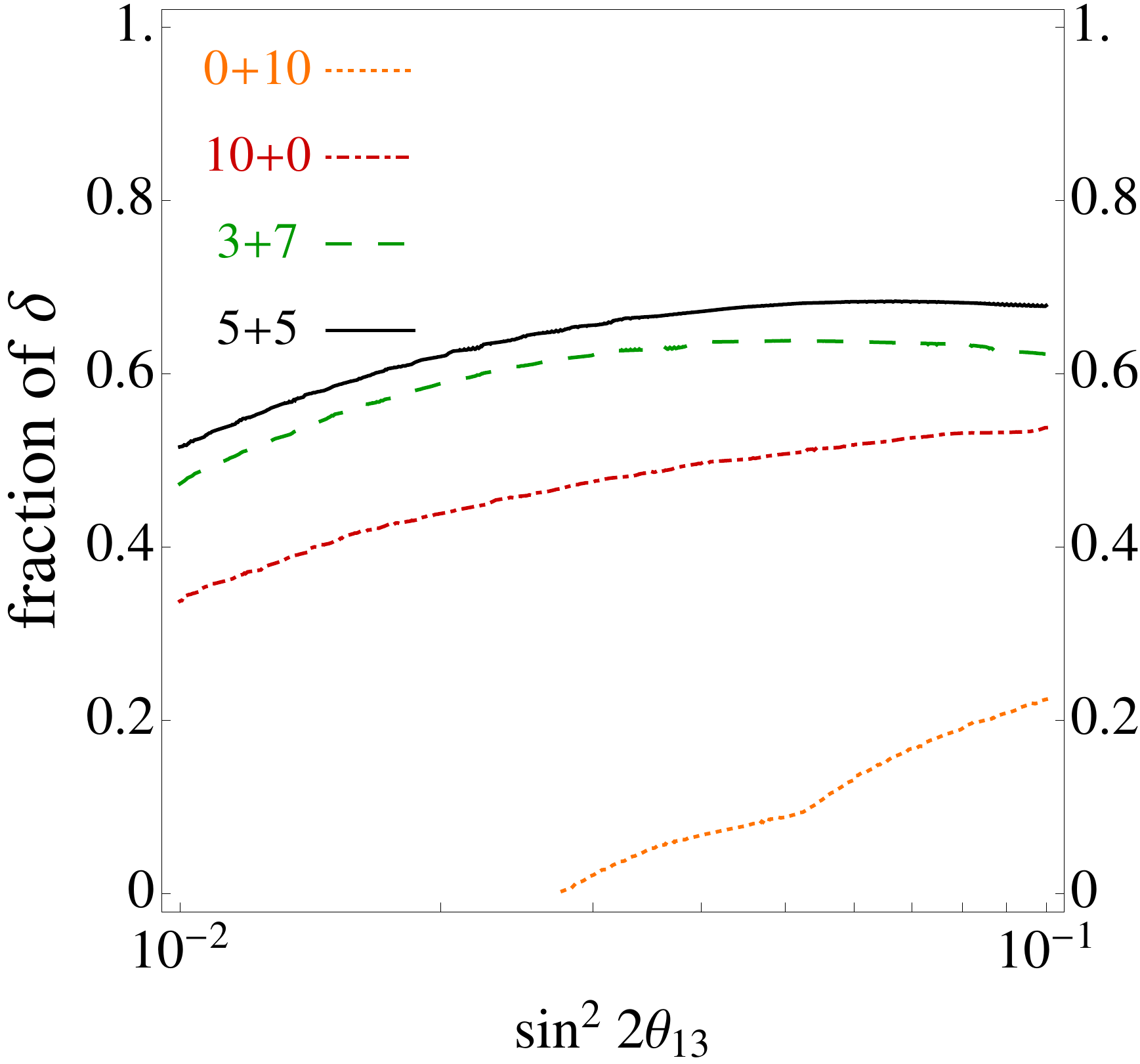}
\caption{Effect of different $\nu$ and $\bar{\nu}$ running times on the CPV discovery potential. The first and second values in the legend refer to the number of years that the experiment is run in $\nu$ and $\bar\nu$ modes, respectively. Results are shown for the LAr detector placed at 2300 km from the source. The line depicts the fraction of possible values of delta for which CPV can be established at a $3\sigma$ CL as a function of the true value of $\sin^22\theta_{13}$ (1 d.o.f.). }
\label{fig:time}
\end{figure}

As stated in Sec.~\ref{sec:beam}, the default configuration in running times for the 130 km setup consists of 2 years in $\nu$ mode and 8 years in $\bar{\nu}$ mode, as in Ref.~\cite{Campagne:2006yx}. This is the optimal configuration at these energies since it gives an approximately equal number of $\nu$ and $\bar\nu$ events at the detector. For longer baselines, however, the wider energy range makes it possible for a measurement on $\delta$ and only few $\bar\nu$ events are needed. Therefore, for the other baselines we have considered a symmetric configuration, that is, 5 years in $\nu$ mode and 5 years in $\bar\nu$ mode. In Fig.~\ref{fig:time} we show how different running times could affect the CPV discovery potential of a given setup. 
  
The reduced flux and cross-section for $\bar{\nu}$ compared to $\nu$ means that for equal running times, the number of events in the $\bar{\nu}$ mode is expected to be much smaller than that obtained in $\nu$ mode. In addition to this, for a normal hierarchy the probability is greatly enhanced for neutrinos while the opposite takes place for antienutrinos. The consequence of both effects is seen from the comparison between the dotted red and dotted blue lines in Fig.~\ref{fig:time}. It can be seen that if the experiment is run only in $\bar{\nu}$ mode, the CPV discovery potential of the facility is quite limited. On the contrary, the results when the experiment is run only in $\nu$ mode are remarkable. This is due to the larger flux and higher cross-section for neutrinos with respect to antineutrinos, as well as to the resonant effect which takes place for neutrinos for a normal hierarchy. Also, the broad peak of the spectrum combined with a detector which has an excellent energy resolution allows to reconstruct the oscillatory pattern and disentangle the CP-violating part. Nevertheless, the fraction of values of $\delta$ for which CP violation can be discovered for $\sin^22\theta_{13}\sim 10^{-1}$ is only slightly above $50\%$. This is considerably improved when the experiment is also run in $\bar\nu$ mode (solid green and black lines). As can be seen from the plot, the best results are obtained when a symmetric configuration in $\nu$ and $\bar\nu$ modes is used, under the assumption of a normal hierarchy. 

For an inverted hierarchy, on the other hand, the resonant effect takes place for antineutrinos while the neutrino event rates are suppressed. However, the larger flux and cross section for the latter are able to partially compensate for this and, as a result, the number of events for both polarities would be more balanced. This generally gives a better result for CPV when the experiment is run for 5 years per polarity with respect to the results shown in Fig.~\ref{fig:time}. Therefore, this is the configuration that has been adopted to simulate all our results for the high energy setups (i.e., for $L>130$ km) that will be presented in the next section, since it gives optimal results for both hierarchies. 

\subsection{Tau detection}
\label{sec:taus}

At a superbeam experiment, the oscillation channel that provides sensitivity to $\delta$ is the $\nu_\mu \rightarrow \nu_e$ channel (and its CP-conjugate). As already explained in the introduction, in order to maximize the oscillation probability all the setups considered here are tuned to their first oscillation peak. In this case, however, not only the $\nu_\mu \rightarrow \nu_e$ signal is maximized, but also the $\nu_\mu\rightarrow \nu_\tau$ oscillation. This oscillation would produce an important background source for the higher energy setups that are considered in this paper, since the cross- section for $\tau$ production at the detector is non-negligible. Therefore, a considerable amount of the $\nu_\mu \rightarrow \nu_\tau$ oscillated neutrinos will eventually produce a $\tau$ at the detector, which can decay into an electron (plus missing energy). These events would constitute a background to the low energy part of the signal since the electron will have lower energy than the $\tau$ which was originally produced. This phenomenon, known as the $\tau$-contamination, has already been studied for the Neutrino Factory~\cite{Donini:2010xk,Indumathi:2009hg}, where the background comes from the $\tau$ decay into muons. It is expected, though, that the non-negligible $\tau$ detection efficiency of a LAr detector, combined with appropriate kinematic cuts, would reduce its impact. In addition, the majority of these events would be affecting the lower energy part of the spectrum, leaving the first oscillation maximum unaffected. A dedicated analysis to study the effect of this background for high energy superbeams is needed, though, with migration matrices to account for the migration of these events to the lower energy part of the spectrum. In the absence of such migration matrices, we have not been able to include it in the present work.  

On the other hand, it may be possible for a LAr detector to detect and identify $\tau^\pm$ leptons, making it possible to observe the $\nu_{\tau}$ and $\bar{\nu}_{\tau}$ appearance channels. However, $\tau$ detection is experimentally very challenging and therefore only of benefit if the additional events produce a significant improvement to the performance of the facility. Up to second order in the perturbative expansion of the probabilities, all terms in the $\nu_\mu\rightarrow \nu_\tau$ oscillation channel also appear in the $\nu_\mu$ disappearance channel, with the exception of the CP violating one. Even in the very optimistic case where a $\tau$ detection efficiency of $50\%$ could be attained, the number of $\tau$ events would be roughly $5\%$ of the $\mu$ events observed at the detector (see Tab.~\ref{tab:taus} in App.~\ref{sec:events}). Therefore, the inclusion of $\nu_\mu \rightarrow \nu_\tau$ events in the analysis would only be helpful if it provided additional information contributing to the CPV discovery potential of the facility. However, the $\delta$-dependence of the number of events for this channel is not as clear as for the $\nu_\mu \rightarrow \nu_e$ channel, since a large fraction of the number of events comes from the leading atmospheric term in the probability (which is CP-conserving). Therefore, we find that the inclusion of the $\nu_{\mu}\rightarrow\nu_{\tau}$ and $\bar{\nu}_{\mu}\rightarrow\bar{\nu}_{\tau}$ channels, even under the very optimistic assumptions of a $50\%$ $\tau$ detection efficiency and a background of $10^{-3}$ of the $\nu_{\mu}$ disappearance events, has a negligible impact on the results.

Finally, this channel may still be useful in searching for the effects of new physics in neutrino oscillations, in particular, non-standard interactions (NSIs)~\cite{Donini:2008wz}. NSIs are flavour-changing interactions due to New Physics occurring at high energies, which may affect neutrinos at production, during propagation, or at the point of detection~\cite{Kopp:2007mi,Kopp:2007ne,Ohlsson:2008gx}. Neutrino oscillation experiments are the perfect hunting ground for such effects (see e.g. Refs.~\cite{GonzalezGarcia:2001mp,Ota:2001pw}).
However, these effects are expected to be strongly suppressed by the scale of the New Physics and the necessary requirement of gauge invariance~\cite{Gavela:2008ra,Antusch:2008tz}. Thus, a powerful neutrino beam experiment would likely be needed to improve them. The best setup for this task is recognised to be the high-energy Neutrino Factory~\cite{Winter:2008eg,Kopp:2008ds,Coloma:2011rq}, but a next-generation superbeam may also be able to improve upon current constraints~\cite{Biggio:2009nt,Biggio:2009kv,Antusch:2010fe}. Concerning NSIs in production and detection, a superbeam would offer the possibility to test different coefficients since the neutrinos are mainly produced from pion and kaon decays~\cite{Antusch:2010fe}, unlike in the Neutrino Factory where neutrinos are exclusively produced through muon decay. Recently, a proposal to build a tau detector for the NuMI beam, MINSIS~\cite{Alonso:2010wu}, has been made in order to search for NSI effects in production and detection processes. Concerning NSIs in propagation, a very interesting possibility to improve present bounds would be to observe the $\nu_{\mu}\to\nu_{\tau}$ and $\bar\nu_{\mu}\to\bar\nu_{\tau}$ channels, the so-called `discovery channels'~\cite{Donini:2008wz}. This has already been proposed for the Neutrino Factory and could also be done at a superbeam experiment.

\section{Comparison between different baselines}
\label{sec:results}

The LAGUNA Design Study offers seven possible baselines for a beam sourced at CERN, ranging from 130 km for the CERN-Fr\'{e}jus setup to 2300 km for CERN to Pyh\"asalmi. The choice of baseline affects  the physics reach in a variety of ways, and impacts on the neutrino spectrum as the energy of the beam is always tuned to match the first oscillation peak. This should be taken into consideration when choosing one detector with respect to another: for instance, at sub-GeV energies WC detectors allow very large fiducial masses to be reached, with excellent energy resolution and background rejection, while for higher energies the LAr technology might be preferred due to the ability to reconstruct very well the non-quasi-elastic events. 

The performance of any neutrino oscillation experiment depends critically on the length of the baseline, since matter effects alter oscillation probabilities relative to the vacuum case (see App.~\ref{sec:probs}). The effect increases with the baseline and/or the density of the medium through which the neutrino propagates. This phenomenon is exploited by long-baseline neutrino oscillation experiments in order to determine the neutrino mass hierarchy (normal or inverted). Resolving the mass hierarchy then leads to increased sensitivities to the other oscillation parameters. 

It should also be taken into account that, even if the number of events at a given baseline scales with $1/L^2$, the cross-section above $\sim1$ GeV increases linearly with the neutrino energy. In addition, while for the high-energy setups ($L>130$ km) the maximum at the flux peak is roughly the same, the width of the peak increases with the energy of the setup, providing a larger integrated flux and also a wider distribution of the events into different energy bins. We have included in App.~\ref{sec:events} the total number of events for all baselines under consideration, for $\sin^22\theta_{13}=0.1$ and several values of $\delta$ (Tab.~\ref{tab:large}). From the total number of events it can be seen that both factors are able to partially compensate for the flux loss. Consequently, in general a better global performance is obtained for the setups with longer baselines. 

Therefore it is vital to compare the performances of each of the LAGUNA baselines (listed in Tab.~\ref{tab:sites}) in addition to that of the detectors, as we do in this section. We present our results in terms of the three observables defined in Sec.~\ref{sec:sim}. The results for the sensitivity to CP violation and the hierarchy will be presented as a function of $\sin^22\theta_{13}$ and $\delta$ at a statistical significance of $3\sigma$ (1 d.o.f). Since the ability to rule out maximal mixing in the atmospheric sector depends mainly on the values of both $\theta_{23}$ and $\theta_{13}$, the results for this observable will be presented as a function of $\sin^22\theta_{13}$ and $\delta \theta_{23} \equiv \theta_{23} - 45^\circ$, at a statistical significance of $3\sigma$ (1 d.o.f). 

Since some of the LAGUNA baselines are very close (630 km and 665 km, as well as 950 km and 1050 km), the results obtained for them are very similar. Therefore we show the results for only one of the baselines in each case, and it should be understood that the results are practically identical to those of the baseline not shown.

We note, for reference, that our results for the WC detector placed at 130 km are consistent with those obtained in Ref.~\cite{Campagne:2006yx}. Similarly, our results for the LAr detector placed at the longest baselines (1570 km and 2300 km) are consistent with those obtained in Ref.~\cite{Rubbia:2010fm}. We stress that the results for the 130~km baseline and the longer ones are obtained for a different accelerator setup and with significantly different neutrino fluxes. Therefore, their comparison should be made with caution and we will present the results in the text separately.

\subsection{CPV discovery potential}
\label{sec:CP}

\begin{figure}[!here]
     \centering
     \subfigure[~LAr]{
          \includegraphics[scale=0.35]{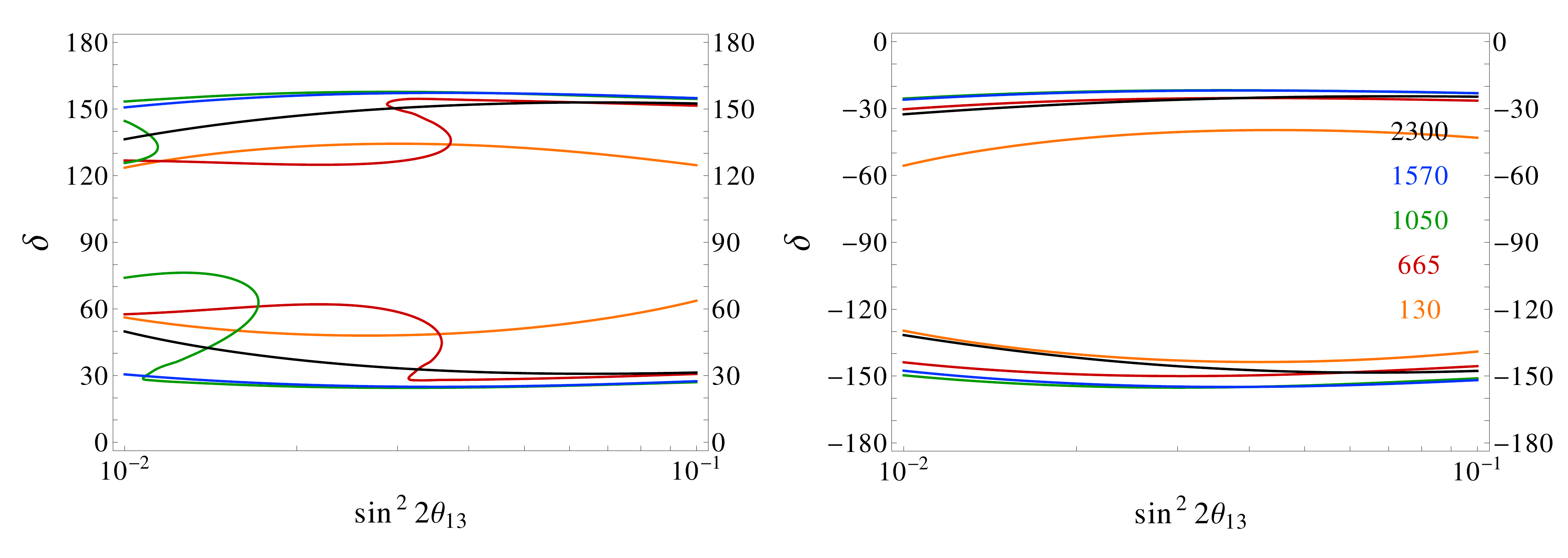}\label{fig:cp_LAr}} 
     \hspace{.3in}
     \subfigure[~WC]{
          \includegraphics[scale=0.35]{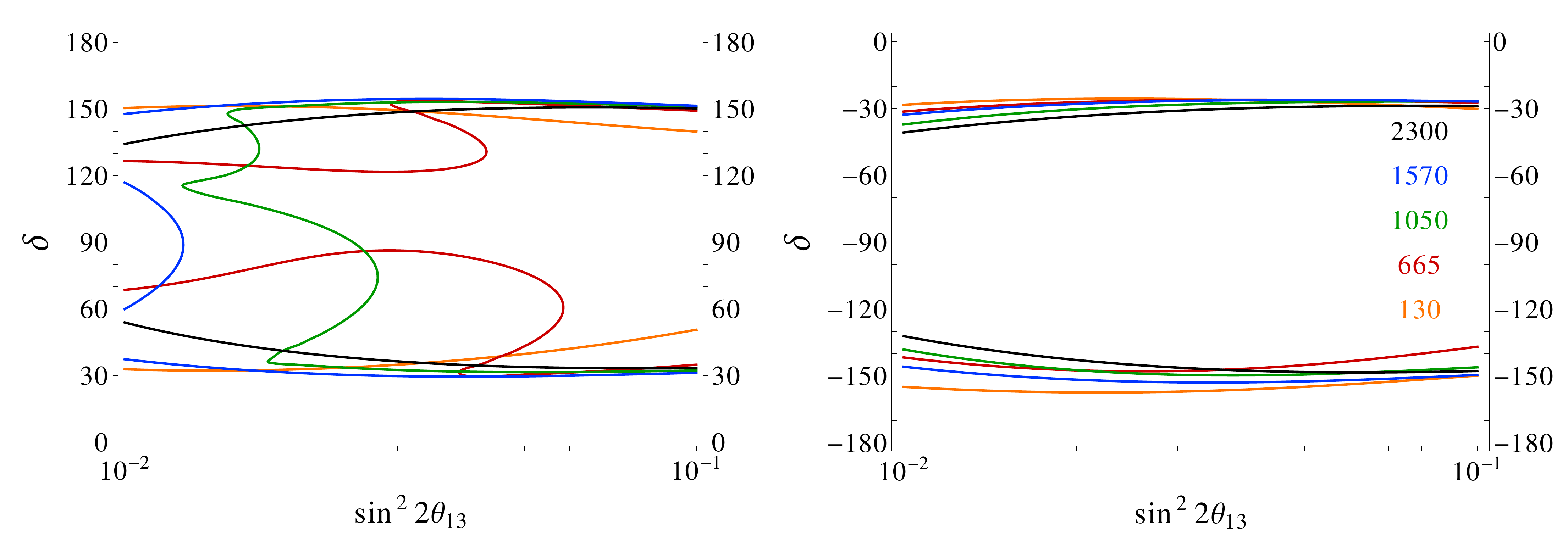}\label{fig:cp_WC}} \\
     \subfigure[~LSc]{
          \includegraphics[scale=0.35]{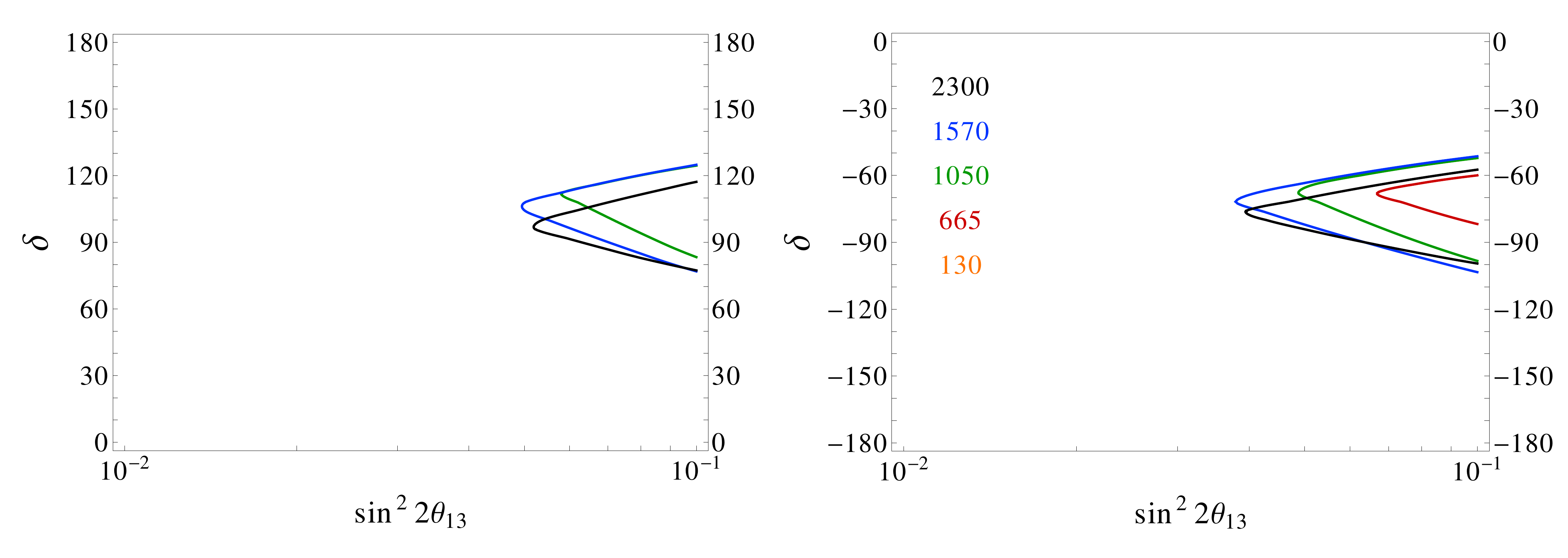}\label{fig:cp_LS_NC}}
     \hspace{.3in}\\
\caption{CPV discovery potential as a function of $\sin^22\theta_{13}$ and $\delta$ for which the CP conservation hypothesis ($\delta=0$ or $180^{\circ}$) can be ruled out at $3\sigma$ (1 d.o.f.).  The different lines correspond to different baselines, as indicated in the legend. Left and right panels show the CPV discovery potential for positive and negative values of $\delta$, respectively. }
\label{fig:cp}
\end{figure}

In Fig.~\ref{fig:cp} we show our results for the $3\sigma$ CPV discovery potential, for the five baselines and three detectors under consideration. The left-hand panels show the results for $\delta<0$ and the right-hand panels for $\delta>0$. In the region to the right of each line, the CP conservation hypothesis ($\delta = 0,\pi$) can be ruled out at $3\sigma$ (1 d.o.f).

For the 130~km baseline, again the WC detector yields the best results due to its much larger mass and excellent performance in the low energy regime. If a LAr is employed instead, the very low energy of the beam limits the statistics at the detector due to the very small cross-sections at these energies. The situation is even worse for the LSc detector due to its smaller fiducial mass and larger background levels.

For $L>130$ km, the baseline dependence is closely related to the presence of stronger matter effects for the longer baselines, although the same reasoning regarding statistics explained at the beginning of this section applies to the 2300 km baseline for the LAr and LSc detectors. In the region where $\delta \sim +90^\circ$, hierarchy degeneracies move to CP conserving values of $\delta$ and therefore the CPV discovery potential of the facility is worsened~\cite{Huber:2002mx} (for an inverted hierarchy, this would apply to $\delta\sim-90^{\circ}$). This can be avoided if the facility is also able to measure the hierarchy and resolve these degeneracies. As a consequence, longer baselines are generally better for this observable in the $\delta>0$ region. 

\begin{figure}[!here]
     \centering
     \subfigure[~LAr]{
          \includegraphics[scale=0.35]{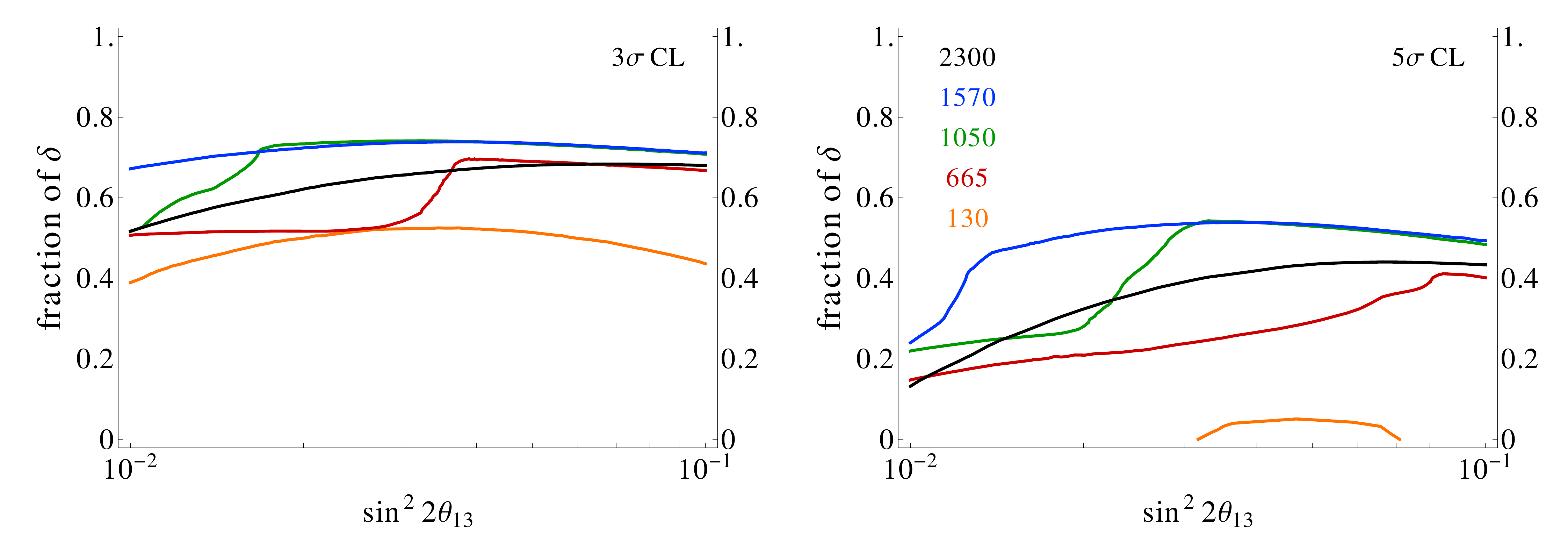}\label{fig:cpfrac_LAr}} 
     \hspace{.3in}
     \subfigure[~WC]{
          \includegraphics[scale=0.35]{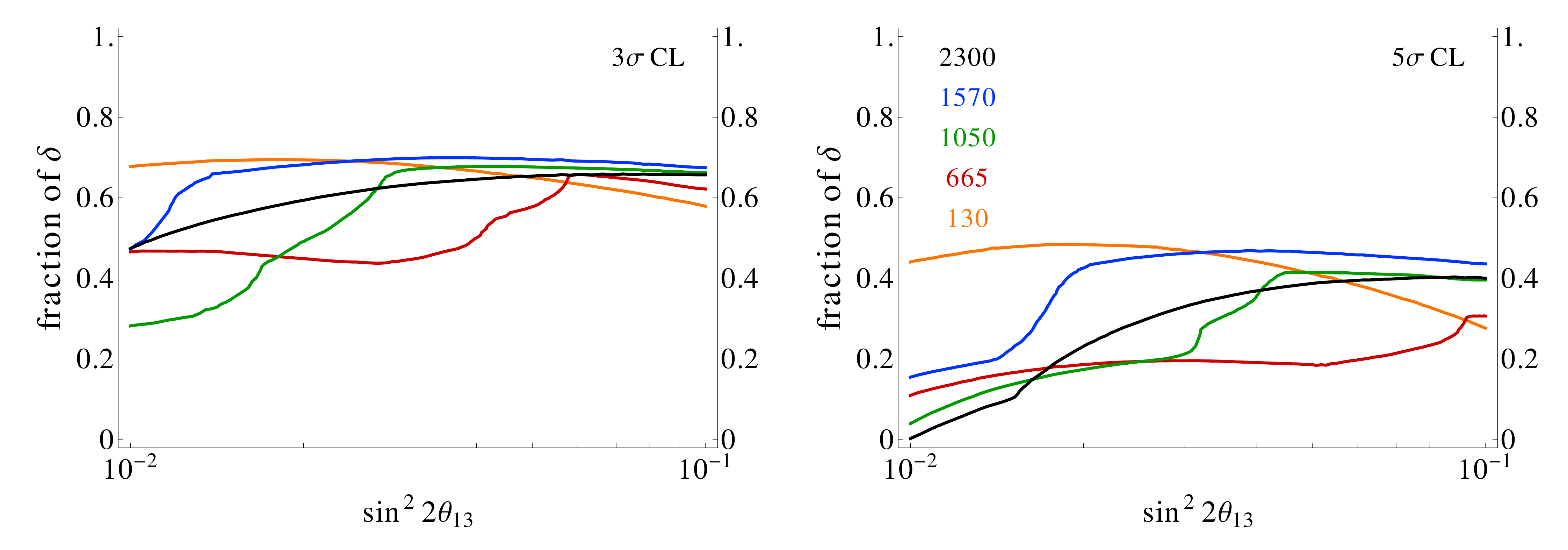}\label{fig:cpfrac_WC}} \\
     \subfigure[~LSc]{
          \includegraphics[scale=0.35]{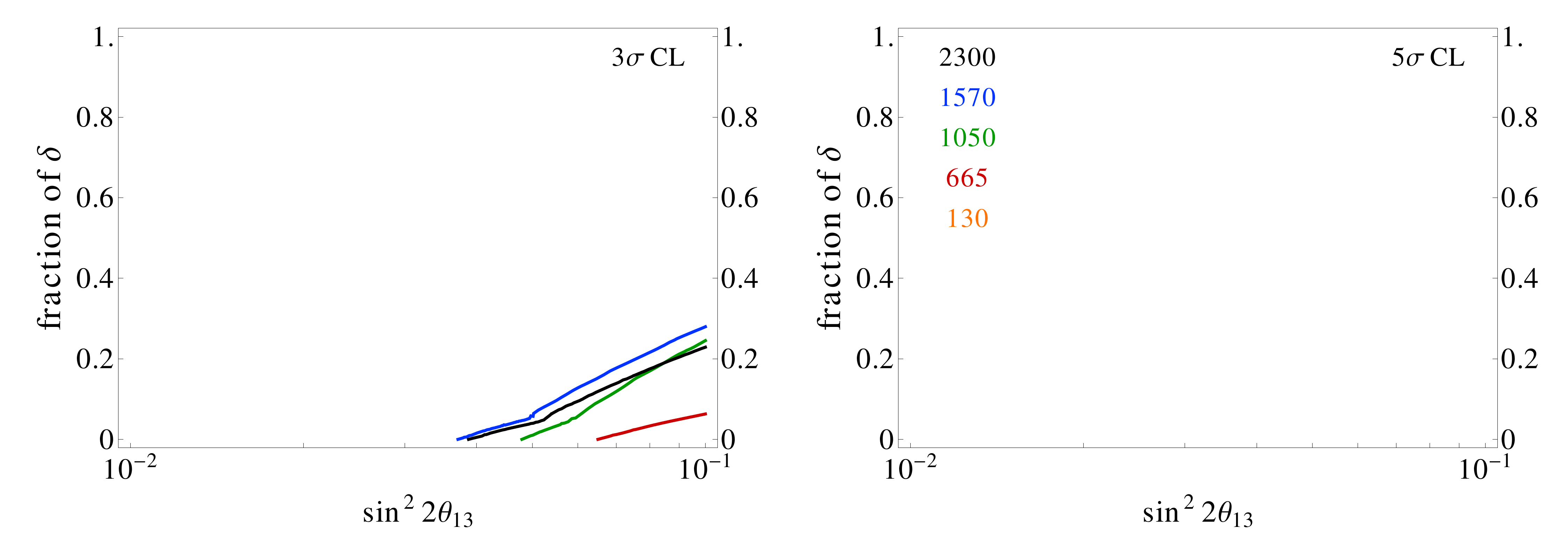}\label{fig:cpfrac_LS_NC}}
     \hspace{.3in}\\
\caption{CPV discovery potential as a function of $\sin^22\theta_{13}$ and $\delta$ for which the CP conservation hypothesis ($\delta=0$ or $180^{\circ}$) can be ruled out at $3\sigma$ 1 d.o.f (left panel) and at 5$\sigma$ 1 d.o.f. (right panel). The different lines correspond to different baselines, as indicated in the legend.}
\label{fig:CPfrac}
\end{figure}

In Fig.~\ref{fig:CPfrac} we show the results for the CP discovery potential as a function of the fraction of possible values of $\delta$ for which CP violation can be established at the $3\sigma$ (left) and $5\sigma$ CL (1 d.o.f.). Results are shown for the five baselines and three detector technologies under consideration. Results are generally better for the LAr detector for all baselines with the exception of the $L=130$ km, for which the WC detector performs best. From the right panels in the figure it can be seen that, for $ \sin^22\theta_{13}\sim 0.1$, all the setups with a LAr detector would be able to discover CP violation for 40\%-50\% of the parameter space, with the exception of the 130 km baseline. If a WC is used instead, the results for the 130 km baseline are comparable to the rest, although slightly worse due to the sign degeneracies, which appear precisely for those values of $\theta_{13}$ and for $\delta>0$.

\subsection{Hierarchy discovery potential}
\label{sec:mass}

\begin{figure}[!here]
     \centering
     \subfigure[~LAr]{
          \includegraphics[scale=0.35]{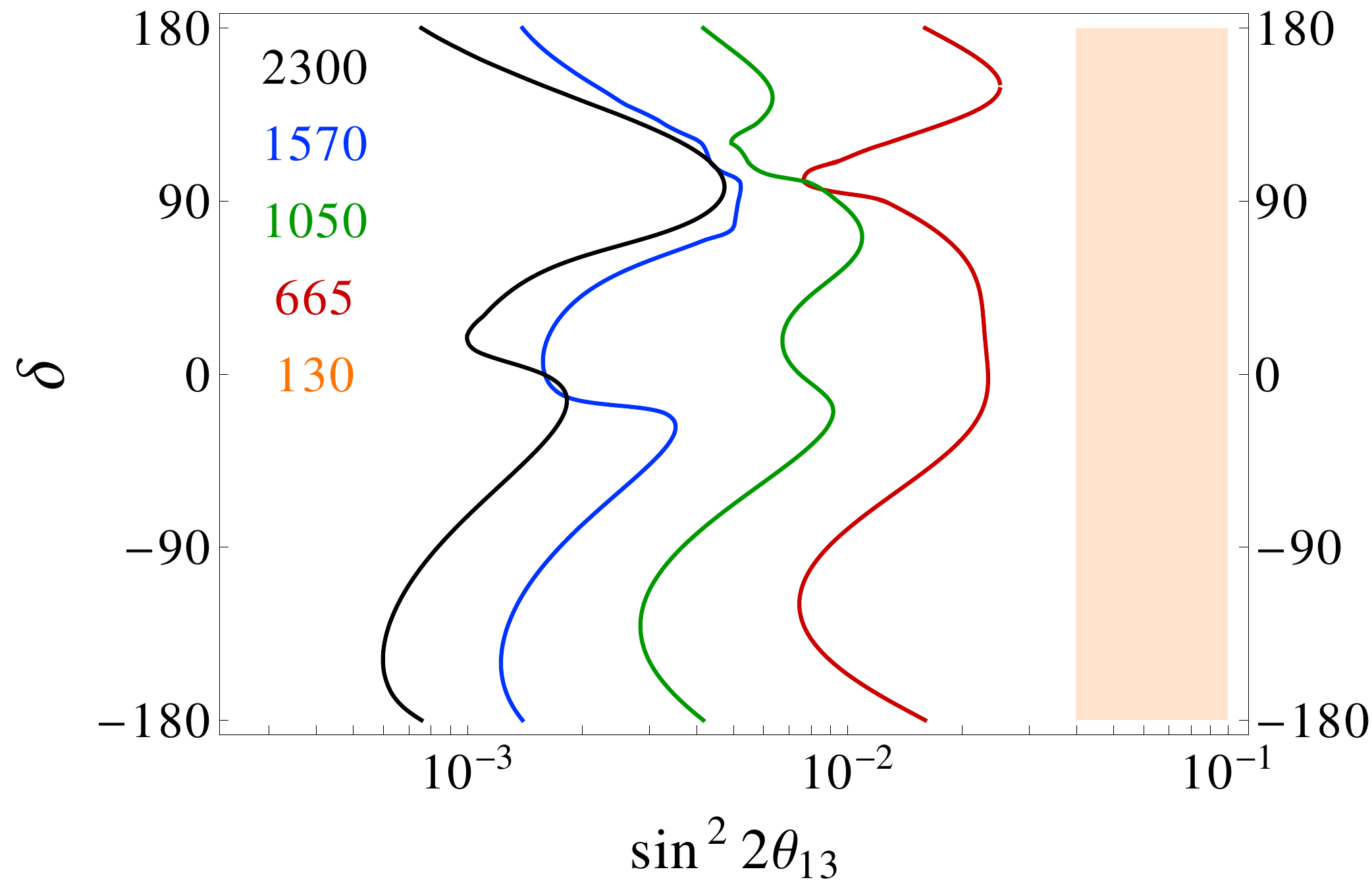}\label{fig:sign_LAr}} 
     \hspace{.3in}
     \subfigure[~WC]{
          \includegraphics[scale=0.35]{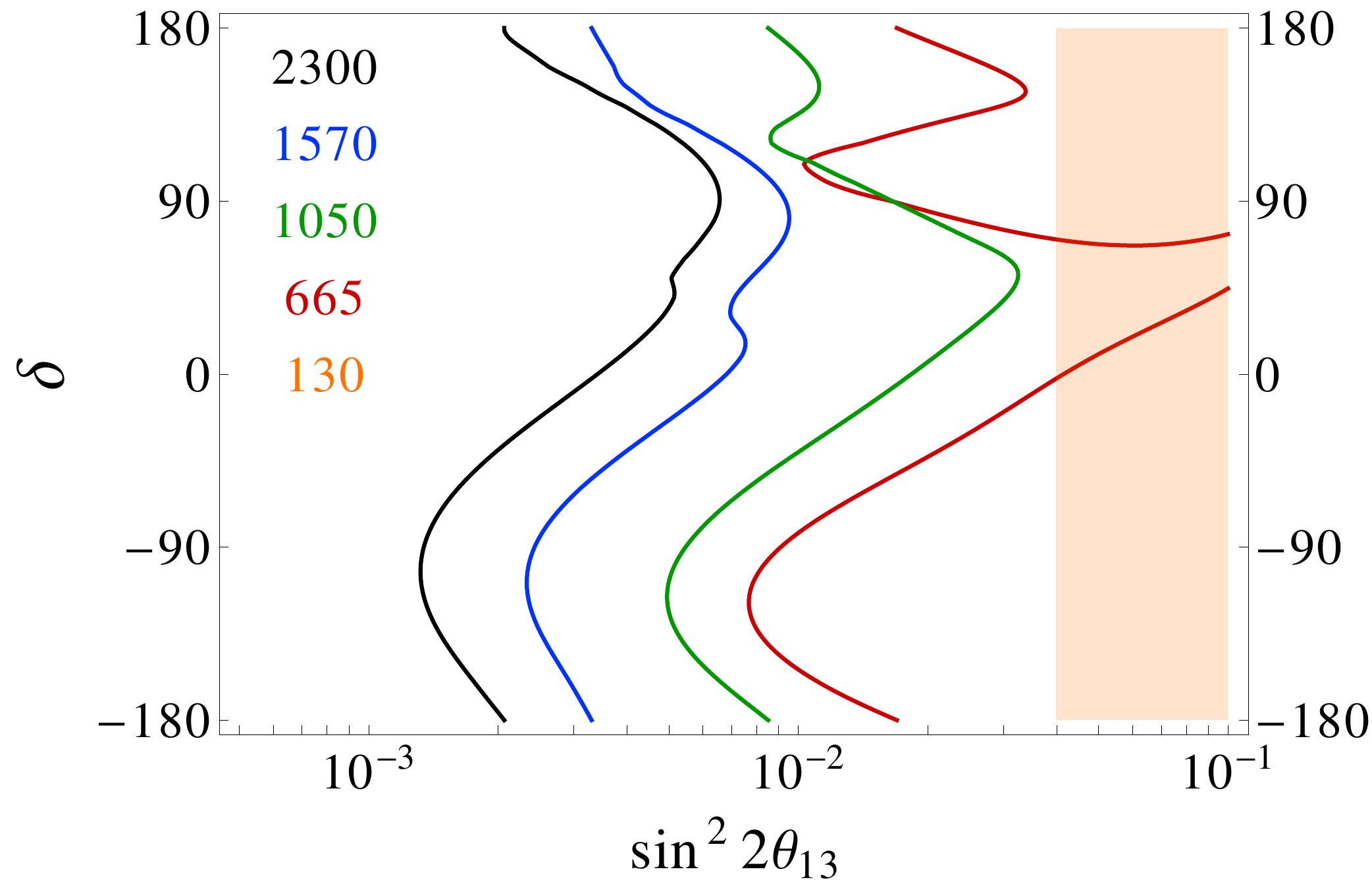}\label{fig:sign_WC}} \\
     \subfigure[~LSc]{
          \includegraphics[scale=0.35]{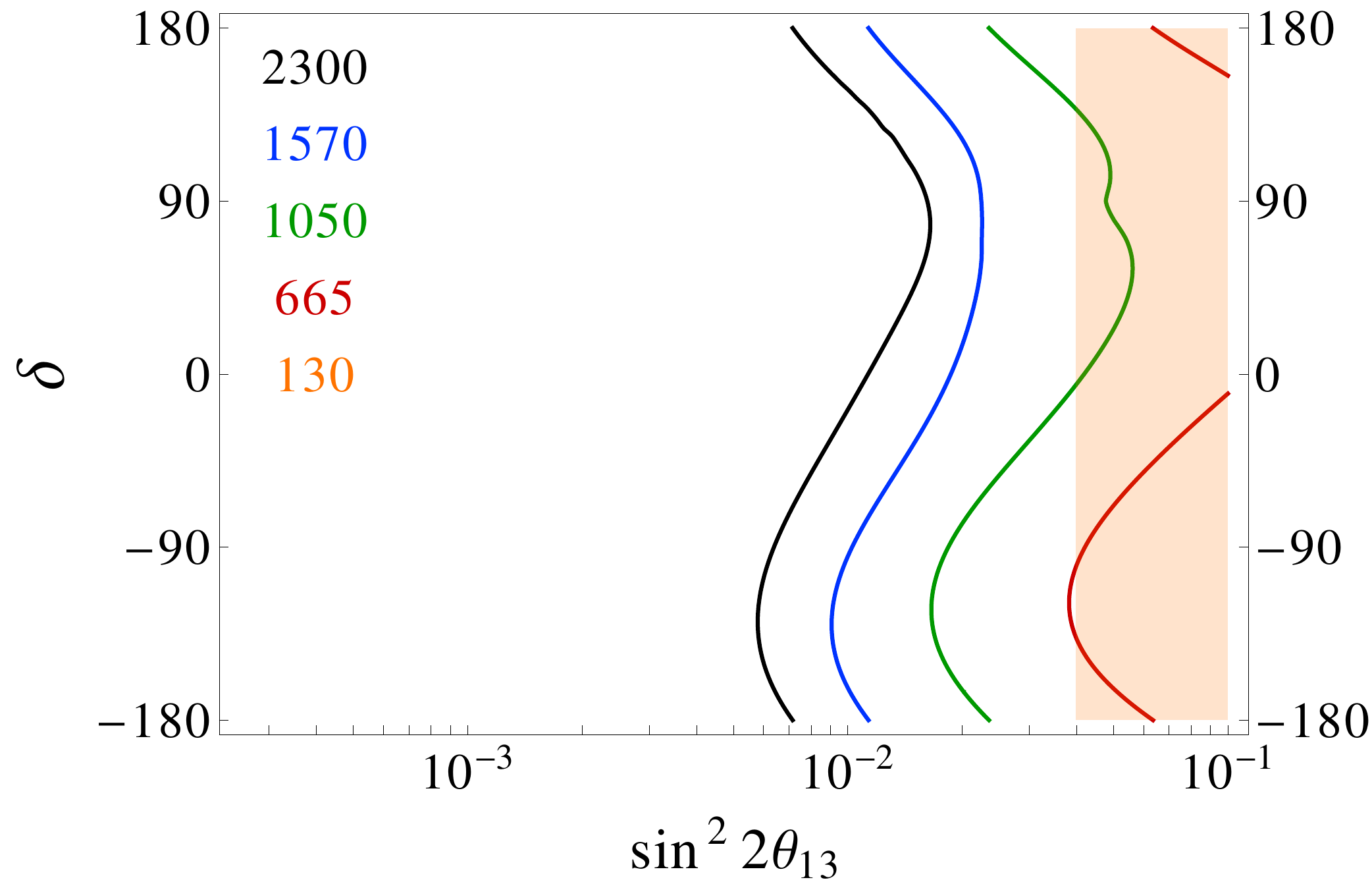}\label{fig:sign_LS_NC}}
     \hspace{.3in}\\
\caption{Hierarchy discovery potential as a function of $\sin^22\theta_{13}$ and $\delta$ for which the wrong hierarchy can be ruled out at $3\sigma$ (1 d.o.f.). A normal hierarchy has been assumed.  The different lines correspond to different baselines, as indicated in the legend. The shaded region corresponds to the $3\sigma$ allowed region at Daya Bay.
\label{fig:sign}}
\end{figure}

In Fig.~\ref{fig:sign}, we present our results for the mass hierarchy discovery potential for the three detectors under consideration and the five baselines under study. Again, a normal hierarchy has been assumed in all cases. We have checked that the results for an inverted hierarchy are very similar to these, but after changing $\delta \rightarrow -\delta$. In these plots we show the results for values of $\sin^{2}2\theta_{13}$ down to $\sim10^{-3}$ and have shaded the region favoured by the Daya Bay results. In the region to the right of each line, the wrong hierarchy can be ruled out at $3\sigma$ (1 d.o.f). Therefore, all the lines which do \emph{not} overlap the shaded region correspond to the setups for which the mass hierarchy can be determined at $3\sigma$ CL for \emph{all} values of $\delta$, assuming the true value of $\theta_{13}$ lies within the interval favoured by Daya Bay at the $3\sigma$ CL. 

Unlike for CPV sensitivity, the choice of baseline is critical in this case. Thus it is always true that the longer the baseline, the better the performance.  This is especially relevant for the LSc detector, which presents a reasonably good performance for this observable, as shown in Fig.~\ref{fig:sign_LS_NC}. It can clearly be seen from all panels in the figure that the best results are obtained when the detector is placed at the 2300 km baseline, regardless of the detector technology. The remaining baselines are ordered according to their length. The worst results are obtained for the shortest baselines: no sensitivity at all is obtained (at $3\sigma$) if the detector is placed at 130 km, whereas in the case of 665 km some sensitivity can be achieved, although not for all values of $\delta$ if combined with a WC or LSc detector.

\subsection{Non-maximal \texorpdfstring{$\theta_{23}$}{} discovery potential}


The sensitivity to $\theta_{23}$ originates primarily from the appearance channel (see App.~\ref{sec:probs}): for `small' values of $\theta_{13}$ ($\sin^22\theta_{13}<10^{-2}$ - not shown in these plots) the sensitivity comes mainly through the solar term and therefore is independent of $\theta_{13}$; as $\theta_{13}$ is increased, the dependence on $\delta\theta_{23}$ which comes through the CP violating term in the appearance channel becomes more relevant. Finally, for $\sin^22\theta_{13}\sim 10^{-1}$, the atmospheric terms in both the appearance and the disappearance channels become relevant and play a very important role. In the presence of matter effects at long baselines, these terms are enhanced and also the neutrino and anti-neutrino probabilities are affected differently. Therefore matter effects are a key factor for this observable, as was the case for the mass hierarchy discovery potential. There is a preference for detecting $\delta\theta_{23}<0$ because some of the terms which are relevant for large $\theta_{13}$ are asymmetric in $\theta_{23}$.

\begin{figure}[here!]
\includegraphics[scale=0.45]{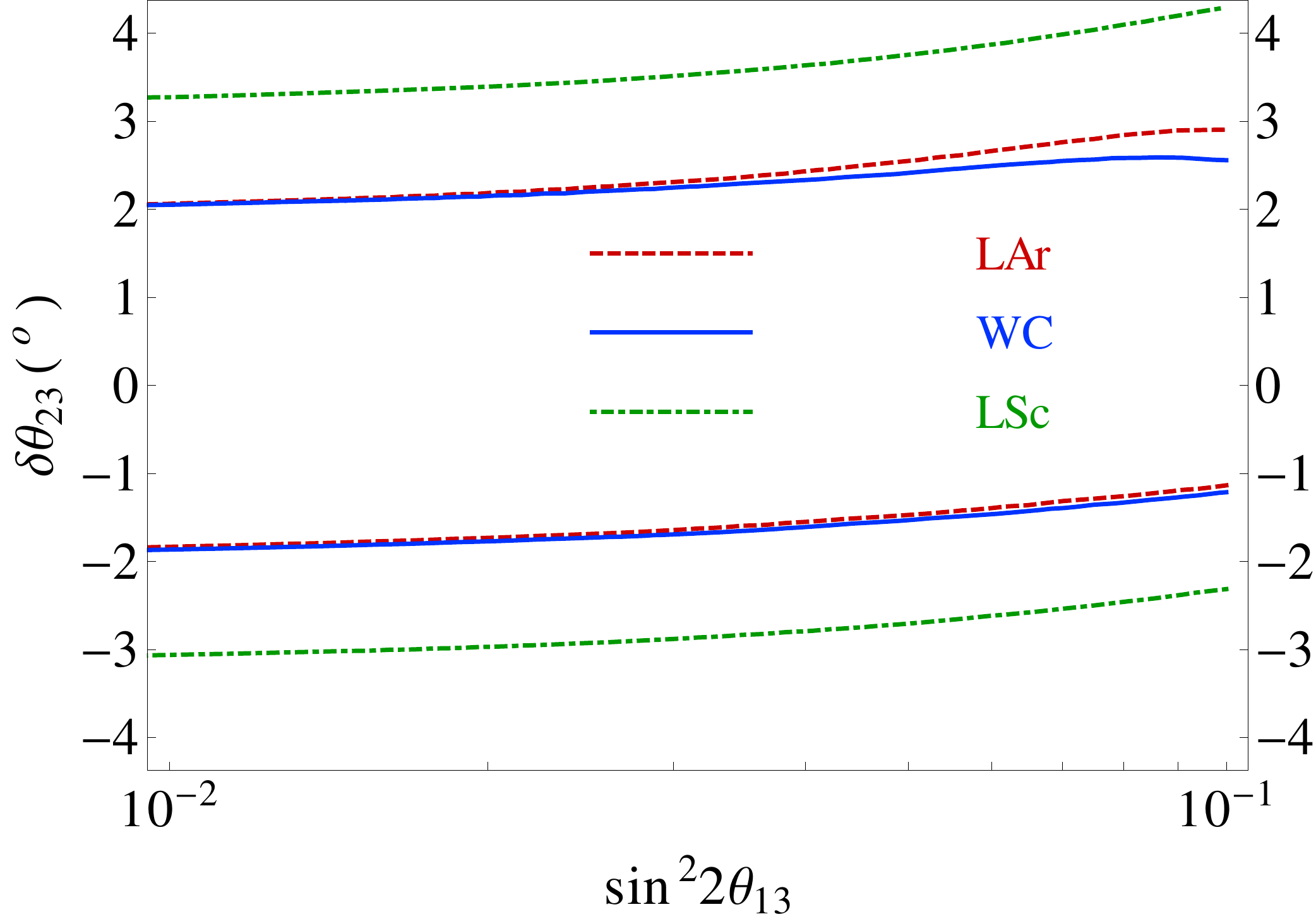}\\
\caption{Non-maximal $\theta_{23}$ discovery potential, as a function of $\sin^22\theta_{13}$ and $\delta\theta_{23} \equiv \theta_{23}- 45^\circ$. In the region enclosed by each line, maximal $\theta_{23}$ cannot be ruled out at a statistical significance of $3\sigma$ (1 d.o.f.). The different lines correspond to different detector technologies, as indicated in the legend. Results are shown for the detector placed at Pyh\"asalmi and for a true normal hierarchy. } 
\label{fig:maximal}
\end{figure}

As long as $L>130$ km, matter effects play a role and our results are practically equal for all the baselines. Mild differences arise between the LAr and WC setups, while the LSc always performs slightly worse due to the larger NC background levels. This can be seen in Fig.~\ref{fig:maximal}, where the results for non-maximal $\theta_{23}$ discovery potential at the Pyh\"asalmi baseline (2300 km) are depicted as a function of  $\sin^{2}2\theta_{13}$ and $\delta\theta_{23} \equiv \theta_{23}- 45^\circ$. In the region enclosed by each line, maximal $\theta_{23}$ cannot be ruled out at a statistical significance of $3\sigma$ (1 d.o.f.), after marginalising over all other parameters. As expected, a mild dependence on $\theta_{13}$ appears for large values of this parameter, while for small values of $\theta_{13}$ the dependence disappears since the sensitivity in this case is achieved through the solar term in the probability.

For $L=130$ km the results are much worse if a LAr or a LSc detector is employed, due to the absence of matter effects. Slightly better results are obtained for the WC due to its much larger mass, although the results are still much worse than for the rest of the baselines. We find that when $\theta_{23}$ is in the second octant (i.e., for $\delta\theta_{23}>0$), a WC detector placed at $L=130$ km could measure deviations from maximal $\theta_{23}$ if $\delta\theta_{23} \gtrsim 3.5^\circ $, while for the rest of the baselines this value is around $\delta\theta_{23} \gtrsim 2^\circ$ for any of the detectors (a similar situation occurs for $\delta\theta_{23}<0$). It is expected that these results would improve if atmospheric data were also included in the analysis (see, for instance, Ref.~\cite{Campagne:2006yx}, where atmospheric data are included in the analysis of a superbeam aiming to a Mton WC detector placed at the Fr\'ejus site).

\section{Exposure and precision}
\label{sec:exposure}

In this section we study how the results depend on the total exposure (power $\times$ time $\times$ detector mass), comparing the longest (2300 km, Pyh\"asalmi) and the shortest (130 km, Fr\'{e}jus) baselines. In this way we can predict how the performance will be affected if any of the three variables are altered from the reference values. For the 2300 km baseline, we have considered only the LAr detector as this is the most promising technology for this baseline, and for the 130 km baseline we again consider the LAr detector and also the WC detector since this is optimal for short baselines. The results for the other baselines are expected to be qualitatively similar. We also present the confidence regions that would be obtained at several confidence levels for the setups listed above, for the best fit from Daya Bay and several values of $\delta$, as an indication of the achievable precision.

\subsection{CPV discovery potential as a function of the exposure}
\label{sec:exposure_cp}

\begin{figure}[hbtp]
\includegraphics[scale=0.45]{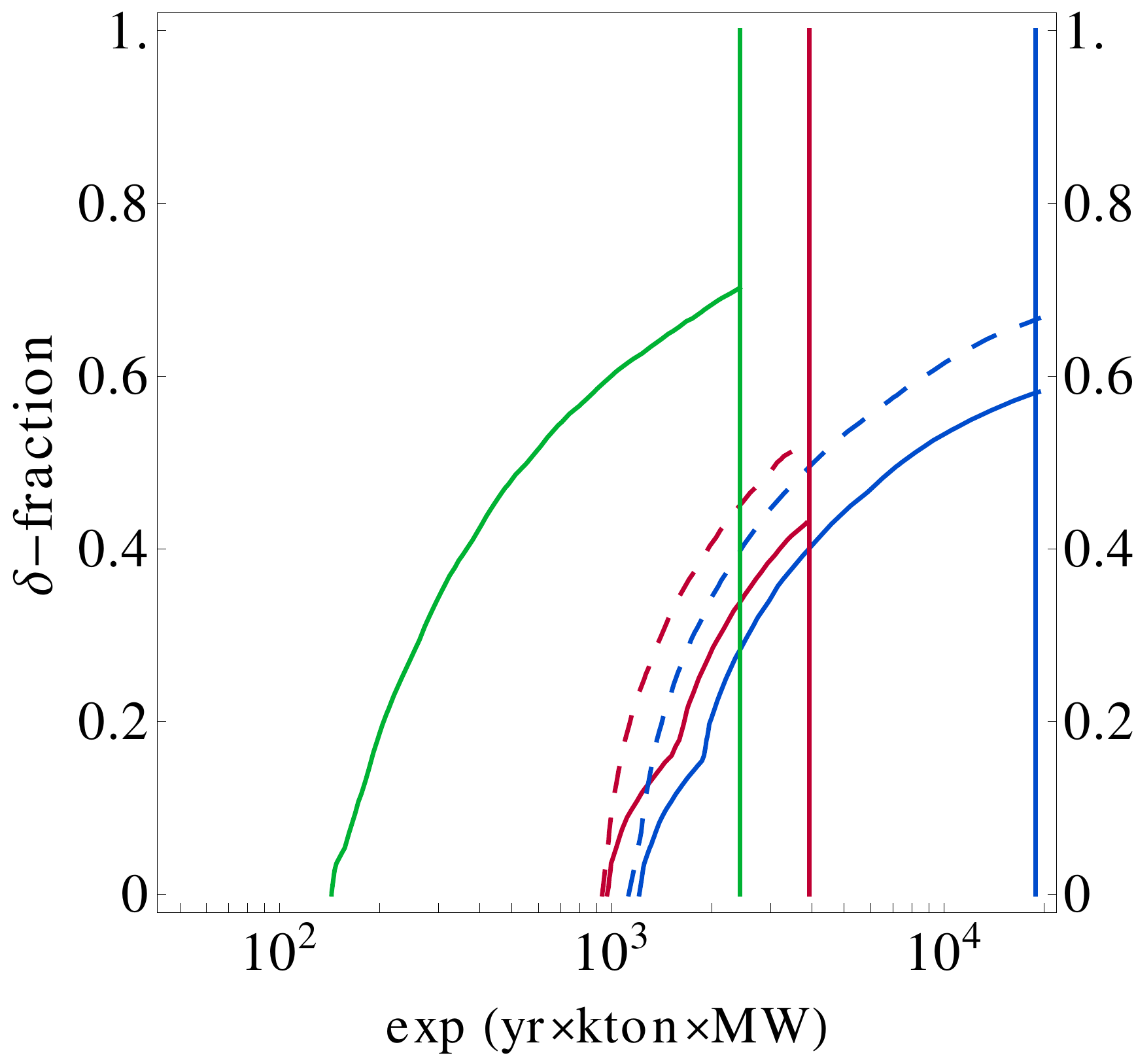}\\
\caption{CPV discovery potential in terms of CP fraction for a true normal hierarchy, as a function of exposure, for LAr at 2300 km (green lines), LAr at 130 km (red lines) and WC at 130 km (blue lines). The solid lines correspond to the situation when the mass hierarchy is still unknown and the dashed lines to the scenario when the mass hierarchy has already been determined externally. The vertical lines correspond to the maximum exposure considered for each of the three setups.} 
\label{fig:cp_exp}
\end{figure}

In Fig.~\ref{fig:cp_exp} the dependence of the CPV discovery potential on the exposure is shown for $\sin^{2}2\theta_{13}=0.092$ (the Daya Bay best-fit value) and as a function of the fraction of possible values of the CP phase for which CPV can be discovered (i.e. $\delta=0,\,180^{\circ}$ can be excluded). Results are shown at the $3\sigma$ CL and for a true normal hierarchy. The green lines correspond to the results for the LAr detector placed at 2300 km, while red and blue lines depict the results for a LAr and WC detector placed at 130 km, respectively. Solid lines correspond to the situation when the mass hierarchy is still unknown and the dashed lines to the scenario when the mass hierarchy has already been determined e.g. by atmospheric neutrino studies or from the combination of INO and/or NO$\nu$A data (see, for instance, Ref.~\cite{Blennow:2012gj}). The vertical lines correspond to the maximum exposure considered for each of the three setups: 10 years$\times$100 kton$\times$2.3 MW for the LAr placed at 2300 km; 10 years$\times$100 kton$\times$4 MW for the LAr placed at 130 km; and 10 years$\times$440 kton$\times$4 MW for the WC placed at 130 km. 

It can be seen how the LAr placed at 2300 km would give the best results with the lower exposure, due to the high energy and wideness of the beam. This setup would be able to establish CP violation at the $3\sigma$ CL in more than $50\%$ of the parameter space for an exposure above $\sim$500 years$\times$kton$\times$MW, which could be achieved, for instance, after 10 years of exposure for a 50 kton LAr detector to a 1 MW beam. On the other hand, for the shortest baseline the exposure needed to have a non-vanishing CPV discovery potential would be much larger. As already explained, this is due to the much smaller neutrino cross-section in the sub-GeV range, which necessarily implies a larger exposure in order to achieve the same results as for higher energy setups. Therefore, in order to achieve the same results the exposure needed for the $L=130$ km baseline needs to be roughly an order of magnitude larger than for the $L=2300$ km setup. Previous knowledge of the mass hierarchy would improve the results by $\sim10\%$ (comparing the solid and dashed lines) if the detector is placed at $L=130$ km, and  we find that it is the total exposure rather than the specific choice of baseline or detector which determines the overall physics reach for this observable.

\subsection{Precision on $\theta_{13}$ and $\delta$}
\label{sec:exposure_precision}

\begin{figure}[!here]
     \centering
     \subfigure[~LAr at 2300 km]{
          \includegraphics[scale=0.26]{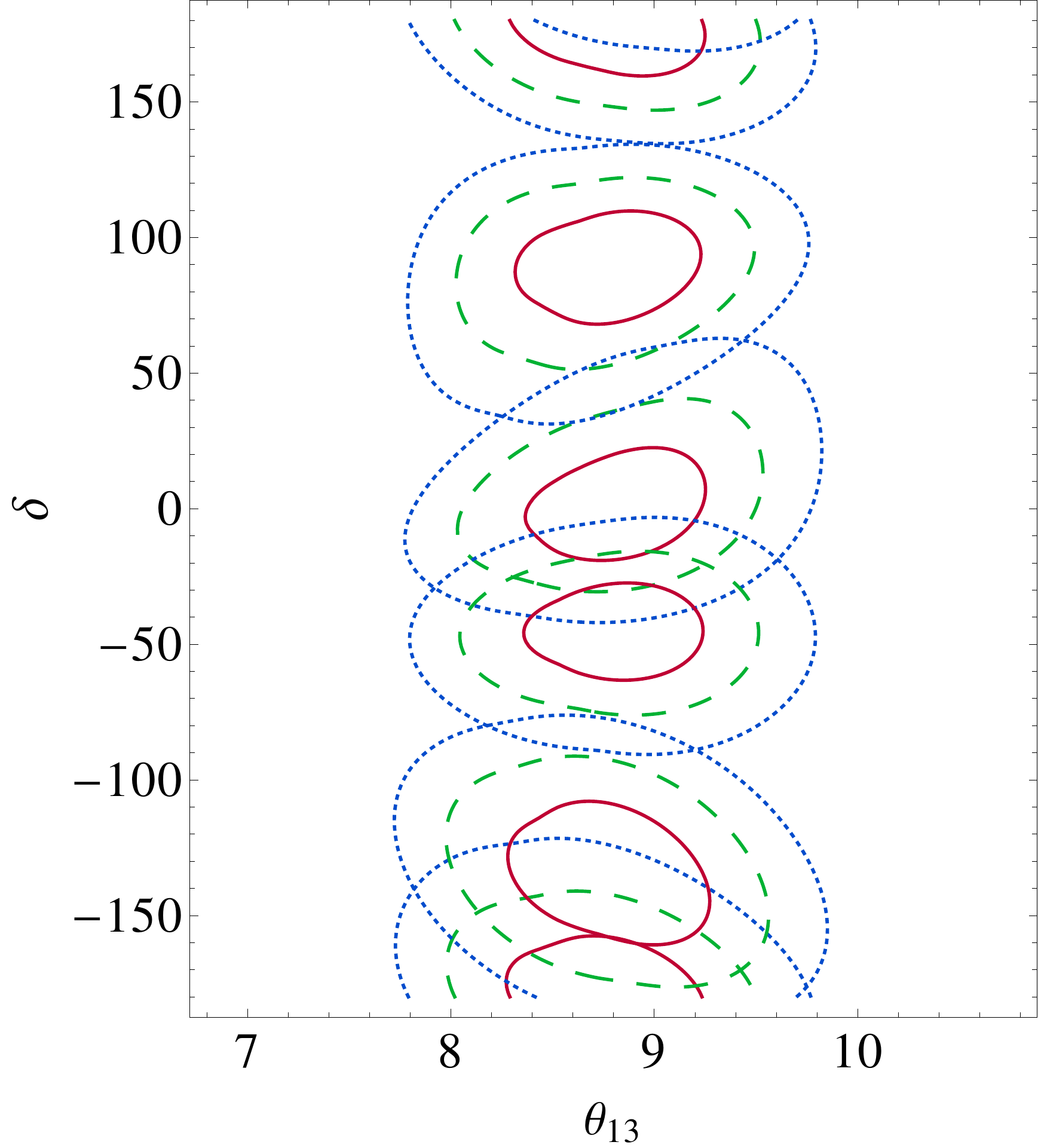}\label{fig:pots_LAr_2300}} 
     \hspace{.3in}
     \subfigure[~LAr at 130 km]{
          \includegraphics[scale=0.26]{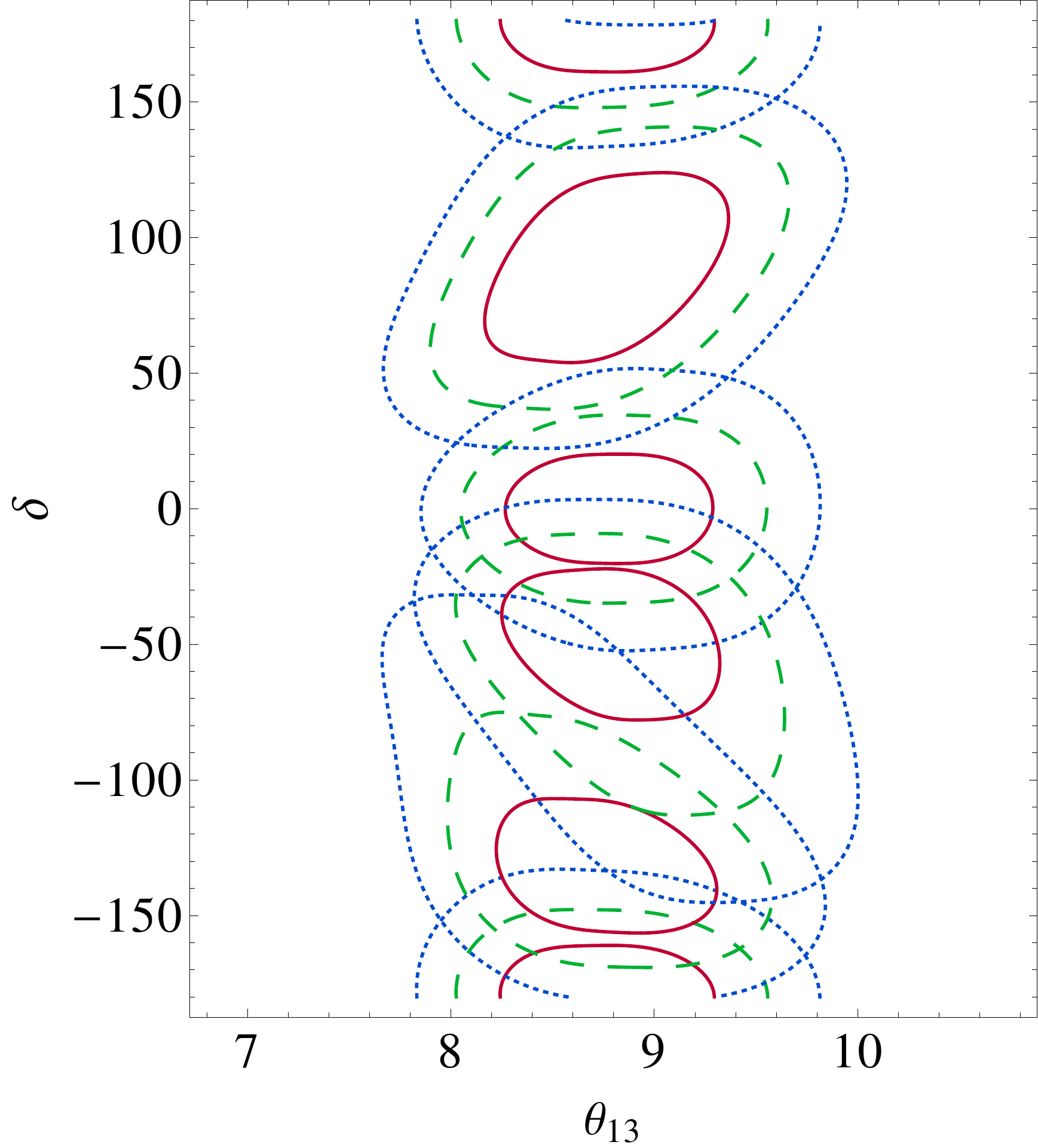}\label{fig:pots_LAr_130}}
      \hspace{.3in}
     \subfigure[~WC at 130 km]{
          \includegraphics[scale=0.26]{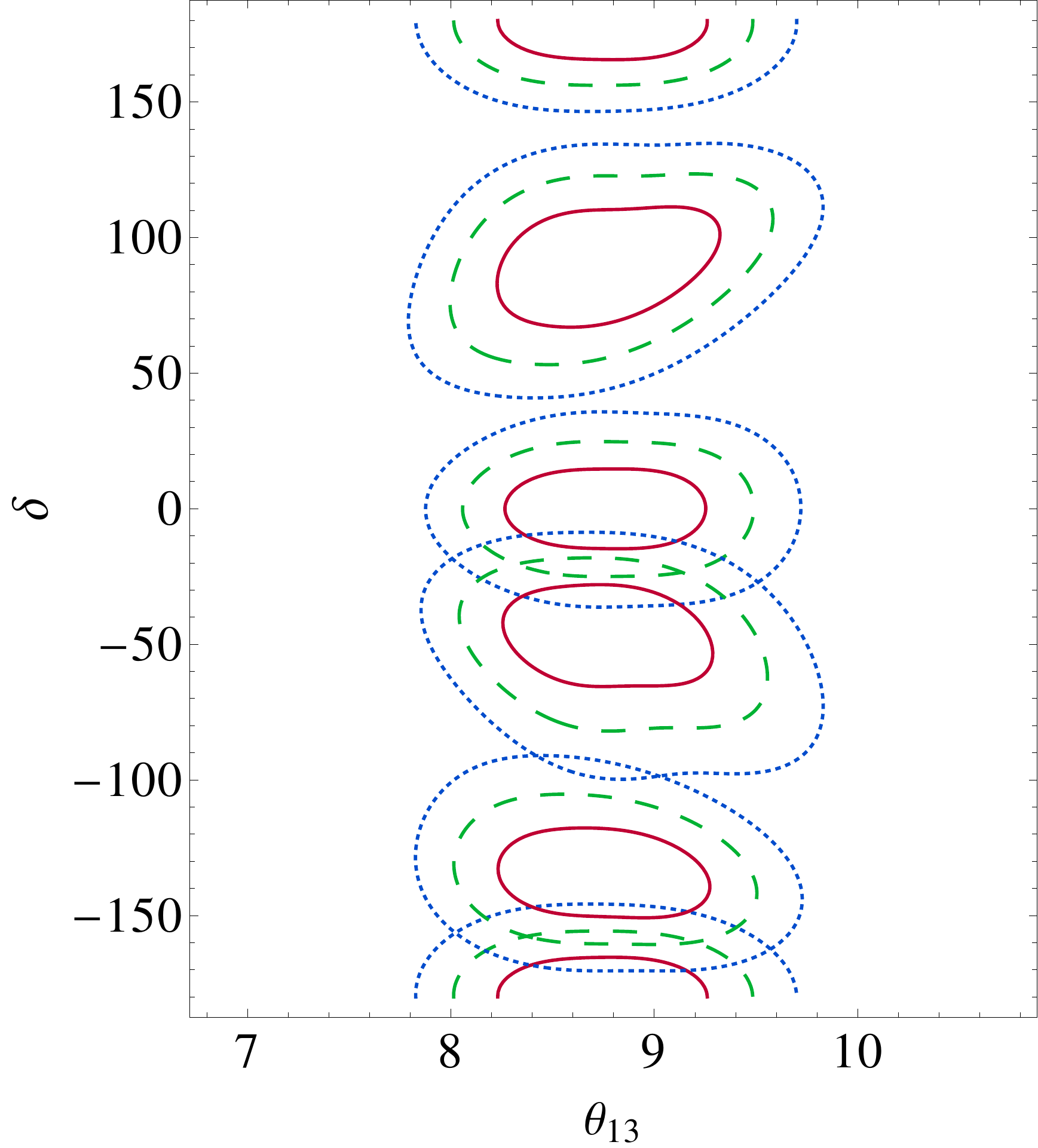}\label{fig:pots_WC_130}} \\
\caption{Confidence regions for 2 d.o.f. corresponding to $1\sigma$ (solid red), $2\sigma$ (dashed green) and $3\sigma$ (dotted blue) in the $\theta_{13}-\delta$ plane obtained by different setups, for a true value of $\theta_{13}=8.83^{\circ}$ (the Daya Bay present best-fit value) and randomly chosen true values of $\delta$.
\label{fig:pots}}
\end{figure}

A recent detailed study of the precision of superbeam and other long-baseline experiments, including the setups presented here for a LAr detector placed at $2300$ km and a WC placed at 130 km from the source, can be found in Ref.~\cite{Coloma:2012wq}. It should be noted, though, that the experimental setup used in that reference for the higher energy option is slightly different from the one considered here (in particular, the beam power is reduced by a factor of 3, for instance). In Fig.~\ref{fig:pots} we show how well each of the three setups would perform in terms of precision for $\theta_{13}$ and $\delta$. Solid (red), dashed (green) and dotted (blue) lines represent the $1,\,2$ and $3\sigma$ contours (2 d.o.f.) in the $\theta_{13}-\delta$ plane, for $\theta_{13}=8.83^{\circ}$ (the Daya Bay best-fit value) and randomly chosen true values of $\delta$. The $\theta_{13}$ precision is similar for all three setups and can be measured to an accuracy of $\sim\pm0.8^{\circ}$ at $3\sigma$ CL by all three setups, independent of the value of $\delta$. For the precision in $\delta$, the LAr detector at 2300 km performs similarly to the WC at 130 km and can measure $\delta$ to $\sim\pm45^{\circ}$. For a LAr detector at 130 km, on the other hand, the precision in $\delta$ is worsened due to the appearance of intrinsic degeneracies around $\delta=\pm 90^\circ$.

\section{Summary and conclusions}
\label{sec:conclusions}

The Daya Bay and RENO collaborations have recently obtained non-zero measurements of $\theta_{13}$ at more than the $5\sigma$ CL~\cite{An:2012eh, collaboration:2012nd} with their best fit values around $\sin^{2}2\theta_{13}\sim 0.09-0.10$. Such a large value of $\theta_{13}$ means that a measurement of the unknown neutrino mixing parameters - in particular the CP phase, $\delta$, and the mass hierarchy - may be within the reach of a next-generation long-baseline experiment. Such an experiment is considered in the European LAGUNA design study, which proposes to construct a multi-kiloton-scale underground neutrino detector within one of seven proposed sites in Europe. This detector could be combined with a superbeam from CERN to create a powerful setup for neutrino oscillation physics, the potential of which we have explored in this work. Such a setup is now the focus of the subsequent LAGUNA-LBNO design study which considers specifically the CERN-Fr\'{e}jus and CERN-Pyh\"asalmi options.

We have studied the capability of each of the proposed baselines, which range in length from 130 km to 2300 km, when combined with each of the options for the detector technology: 100 kton of Liquid Argon, 50 kton of Liquid Scintillator or 440 kton of Water \v{C}erenkov. For the shortest baseline we considered a beam configuration corresponding to $5.6\times10^{22}$ protons on target (PoT) per year with a proton energy of 4.5 GeV. For all other baselines the beam flux corresponded to $3\times10^{21}$ PoT per year with 50 GeV protons. This should be taken into account when comparing the results for $L=130$ km to those obtained for the rest of the baselines, since they correspond to very different neutrino beams, which would be produced with a different accelerator complex and a different number of PoTs and proton energy. 

In the first half of this work we studied in detail several factors related to the beam and detector parameters that could have a sizeable impact on the physics reach of the experiment. We studied the effects of the systematic errors, the intrinsic beam background, the NC background, the ratio of $\nu:\bar{\nu}$ running times and the possibility of tau detection. For the values of $\theta_{13}$ favoured by the Daya Bay measurement, we found that the systematic error on the signal has some effect on the CPV discovery potential, as does the level of the NC background. The measurement of the mass hierarchy is not affected significantly since the length of the baseline and the intensity of the matter effects are the relevant factors for this measurement. For long baselines, we find that equal $\nu$ and $\bar{\nu}$ running times give better results than asymmetric configurations. Tau detection is not only very challenging, but does not significantly improve the sensitivity to oscillation parameters for any of our setups.  

In the second half of our work, we directly compared the performances of each of the baselines and detectors, in terms of the discovery potentials for CP violation, the mass hierarchy, and non-maximal mixing in the atmospheric sector. In general, we find an overall better performance as $L$ is increased, in spite of the strong $1/L^2$ decrease expected in the number of events due to the beam divergence. This is due to several reasons which partially compensate for the flux loss: (1) longer baselines imply higher neutrino energies (to match the first oscillation peak) and so larger cross-sections are obtained at the detector; (2) for the $L>130$ km setups, even though the maximum flux is similar, the width of the peak increases with the energy of the beam, providing a wider distribution of events into the different energy bins; and (3) better sensitivity to the mass hierarchy at longer baselines generally translates into better sensitivity to CP violation due to the lifting of degeneracies. However, for very long baselines (above 1000 km) these are not able to compensate the quadratic suppression with the baseline and the number of events at the detector is lower for these baselines. Finally, it should be noted that even though matter effects produce a resonance in the neutrino oscillation probability for NH, the effect is the opposite in the antineutrino channel (the situation would be precisely the opposite for an IH). This has important consequences for the observation of CPV, for which a balanced number of events for both polarities is generally needed to solve the intrinsic degeneracy. 

For CPV discovery, for the values of $\theta_{13}$ favoured by Daya Bay and RENO, the results obtained with either a WC or LAr detector at any of the baselines other than 130 km are very similar and can cover $60\%-70\%$ of the parameter space, depending on the baseline. A WC detector placed at 130 km shows an excellent performance for smaller values of $\theta_{13}$, but this performance is weaker around $\sin^{2}2\theta_{13}\sim10^{-1}$ due to the sign degeneracy, which for this baseline appears precisely in this region for $\delta>0$. A LAr detector at 130 km presents a worse CPV discovery potential, due to its smaller mass which is not able to compensate for the small neutrino cross-sections at the detector in the sub-GeV range. The performance of a LSc detector, at all baselines, is severely limited by the level of NC background which has been assumed for this detector and also by its relatively small mass and low efficiency. 

For hierarchy discovery the results essentially depend only on the length of the baseline. In spite of its strong performance for CP violation, the 130 km baseline has no sensitivity to the hierarchy (at $3\sigma$ CL) even with the WC detector, due to the absence of matter effects. In the $1\sigma$ range favoured by the Daya Bay results, $\sin^{2}2\theta_{13}>0.071$, the 665 km baseline does not cover all values of $\delta$ if combined with a WC or LSc detector, although with a LAr detector it does. All the remaining baselines have full $3\sigma$ coverage with any of the three detectors.

We have also studied the discovery potential for non-maximal mixing in the atmospheric sector. The results for this observable have a mild dependence on $\theta_{13}$. We find that all baselines with $L>130$ km perform similarly: any setup with either a LAr and a WC would be able to distinguish a non-maximal $\theta_{23}$ for $\delta\theta_{23} \sim2.2^{\circ}$ to $\sim3.2^{\circ}$ if $\theta_{23}$ lies in the second octant ($\theta_{23}>45^{\circ}$) and from $\delta\theta_{23} \sim -1.2^{\circ}$ to $-2.2^\circ$ if it lies in the first octant ($\theta_{23}<45^{\circ}$), depending on the value of $\theta_{13}$. The results for the LSc detector would be roughly $1^\circ$ worse due to the larger level of NC backgrounds. The setups with $L=130$ km do not have any sensitivity at all to this observable due to the absence of matter effects. The only exception is for the WC due to its huge fiducial mass, although the results are worse than for the rest of the baselines under consideration. 

In the last section we showed how the performance of the superbeam setup varies with the exposure (running time $\times$ detector mass $\times$ power), comparing the results obtained from the longest baseline (2300 km) with a LAr detector and the shortest baseline (130 km) with either a LAr or WC detector, for the best-fit value of $\sin^{2}2\theta_{13}$ obtained by the Daya Bay experiment. In terms of CPV discovery potential, we find that it is the total exposure rather than the specific choice of baseline or detector which determines the overall sensitivity. With a LAr placed at 2300 km from the source, CPV could be discovered in more than $50\%$ of the parameter space after an exposure of $\sim500$ kton$\times$yr$\times$MW. We find that, in order to obtain the same reach in terms of CPV discovery potential, roughly an order of magnitude larger exposure would be needed in the case of $L=130$ km with respect to the $L=2300$ km case. If the mass hierarchy is previously determined e.g. by atmospheric neutrino experiments, then the sensitivity of the 130 km setup is increased by roughly $10\%$ relative to the case when the mass hierarchy is unknown, so that at the maximum exposure a WC detector can make a $3\sigma$ discovery of CP violation for $\sim 70\%$ of the possible values of $\delta$, while for the LAr detector this value would be around $\sim 40\%$. In terms of the precision which can be obtained on a measurement of $\theta_{13}$ and $\delta$, all three setups show a similar performance for $\theta_{13}$ in terms of precision and can make a measurement with a $3\sigma$ error of $\sim\pm0.8^{\circ}$. However, for the achievable precision in $\delta$, the LAr detector at 2300 km and WC detector at 130 km perform better than the LAr detector at 130 km, with a $3\sigma$ precision of $\sim\pm45^{\circ}$.

To conclude, the recent measurement of a large non-zero $\theta_{13}$ means that determining whether there is CP violation in the leptonic sector, and whether the neutrino mass hierarchy is normal or inverted, may be achievable in the relatively near future with a next-generation superbeam experiment.  The fact that $\theta_{13}$ is large means that the emphasis for future long-baseline experiments is no longer on maximising statistics and minimising backgrounds, but shifts instead to minimising the systematic errors. The technology needed to produce a superbeam experiment is expected to be at hand in the near future and thus we believe that a CERN-based superbeam experiment, aimed to any of the LAGUNA sites, could provide an excellent opportunity to address both of these questions and should be explored further.

\section*{NOTE ADDED}

Even though the results presented here are embedded in the context of the LAGUNA Design Study, the collaboration is not responsible for any of the results presented in this work. It should also be noted that the conclusions extracted from this work may not be applicable to the subsequent LAGUNA-LBNO Design Study. In particular, some technical details have changed in LAGUNA-LBNO with respect to the ones used here, such as the beam power for the high energy options and the fiducial mass for the Water \v{C}erenkov detector. Such changes could have relevant consequences and change the main conclusions extracted from this work. Further changes are also expected if migration matrices are included to simulate the detector responses, something which will be addressed in the context of LAGUNA-LBNO. This should be taken into account when comparing the results presented in this work to any future output from the LAGUNA-LBNO collaboration. 

\vspace{1cm}

\section*{Acknowledgments}

We would like to thank A.~Donini, E.~Fern\'andez-Mart\'inez and A.~Rubbia for illuminating discussions. We would like to thank A.~Longhin for providing the fluxes for the seven baselines, and B.~Choudhary, L.~Esposito, D.~Hellgartner, R.~Mollenberg, A.~Rubbia, M.~Wurm and L.~Whitehead for providing the migration matrices and experimental details that have been used to simulate the detectors responses. We would also like to thank S.~Agarwalla and A.~Rubbia for pointing out an error in the flux normalization. 

This work was funded by the European Community under the European Commission Framework Programme 7 Design Study LAGUNA (Project Number 212343). The EC is not liable for any use that may be made of the information contained herein. 

PC has been supported by Comunidad Aut\'onoma de Madrid. PC also acknowledges financial support through project FPA2009-09017 (DGI del MCyT, Spain), project HEPHACOS S2009/ESP-1473 (Comunidad Aut\'onoma de Madrid) and from the Spanish Government under the Consolider-Ingenio 2010 programme CUP, “Canfranc Underground Physics”, project number CSD00C-08-44022. PC would like to thank the Institute for Particle Physics Phenomenology (Durham, UK) for hospitality during the early stages of this work. TL acknowledges financial support from the European Union under the European Commission FP7 Research Infrastructure Design Studies EUROnu (Grant Agreement No. 212372 FP7-INFRA-2007-1).

\appendix

\section{Oscillation probabilities}
\label{sec:probs}

The phenomenological discussions in this paper are based on the oscillation probabilities for the $\nu_{\mu}\to\nu_{e}$ and $\bar{\nu}_{\mu}\to\bar{\nu}_{e}$ channels, including matter effects.
For this reason, in this appendix we present the corresponding oscillation probabilities in vacuum and in matter. The number of oscillated $\nu_\mu \rightarrow \nu_e$ events is given in App.~\ref{sec:events}. 

 The oscillation probabilities for the $\nu_\mu \rightarrow \nu_e $ and $\bar\nu_\mu \rightarrow \bar\nu_e  $ channels are identical to those of their T-conjugate channels, the $\nu_{e}\to\nu_{\mu}$ and $\bar{\nu}_{e}\to\bar{\nu}_{\mu}$ channels (the so-called `golden channels'~\cite{Cervera:2000kp}), with the exchange $\delta\to-\delta$. The exact forms can be derived analytically, as in Ref.~\cite{Zaglauer:1988gz}, but these are very complicated and the physics is not easily extracted from them. Instead, a precise but far simpler approximation can be obtained by performing an expansion in the parameters $\theta_{13},\,\Delta_{12}/\Delta_{23},\,\Delta_{12}/A$ and $\Delta_{12}L$ where $\Delta_{ij}=\Delta m_{ij}^{2}/2E$ and $A$ is the matter parameter, $A=\sqrt{2}G_{F}n_{e}$, where $n_{e}$ is the electron density. The result is Eq. (16) of Ref.~\cite{Cervera:2000kp} (we have exchanged $\delta\to-\delta$),

\begin{subequations}
\begin{eqnarray}
P_{\mu e}^{\mathrm{mat}} & = & 
\sin^2 2 \theta_{13}\sin^2 \theta_{23}\left ( \frac{ \Delta_{13}}{ \tilde B_\mp } \right )^2
   \, \sin^2 \left( \frac{ B_\mp \, L}{2} \right) \label{eq:mat_atm} \\ 
&+& 
\cos^2 \theta_{23} \sin^2 2 \theta_{12} \left( \frac{ \Delta_{12} }{A} \right )^2 
   \, \sin^2 \left( \frac{A \, L}{2} \right ) \label{eq:mat_sol}\\
& + &
\sin 2 \theta_{13} \cos\theta_{13} \sin 2 \theta_{12} \sin 2 \theta_{23} \frac{ \Delta_{12} }{A} \, \frac{ \Delta_{13}}{ B_\mp } 
   \, \sin \left( \frac{ A L}{2}\right) 
   \, \sin \left( \frac{B_{\mp} L}{2}\right) 
   \, \cos \left( \mp \delta - \frac{ \Delta_{13} \, L}{2} \right ),\nonumber \\ \label{eq:mat_cp} 
\end{eqnarray}
\label{eq:matter}
\end{subequations}where $ B_\mp \equiv |A \mp \Delta_{13}|$ and the upper (lower) signs apply to neutrinos (anti-neutrinos). This is to be compared with the probabilities in vacuum ($A=0$), as given by Eq. (7) of Ref.~\cite{Cervera:2000kp},

\begin{subequations}
\begin{eqnarray}
P_{\mu e}^{\mathrm{vac}} & = & 
\sin^2 2 \theta_{13} \sin^2\theta_{23} \sin^2 \left ( \frac{\Delta_{13} \, L}{2} \right ) \label{eq:vac_atm} \\
& + & \cos^2\theta_{23} \sin^2 2 \theta_{12} \sin^2 \left( \frac{ \Delta_{12} \, L}{2} \right ) \label{eq:vac_sol}\\
& + & \sin 2 \theta_{13} \cos\theta_{13} \sin 2 \theta_{12} \sin 2 \theta_{23}
\frac{ \Delta_{12} \, L}{2} \sin \left ( \frac{ \Delta_{13} \, L}{2} \right ) \cos \left ( \mp \delta - \frac{ \Delta_{13} \, L}{2} \right ).  \label{eq:vac_cp}
\end{eqnarray}
\label{eq:vacuum}
\end{subequations}
Three terms are involved in each expression: the first term, \eqref{eq:mat_atm} and \eqref{eq:vac_atm}, is known as the atmospheric term as it depends only on the atmospheric mixing parameters but not on the solar ones. Similarly, the second term, \eqref{eq:mat_sol} and \eqref{eq:vac_sol}, is know as the solar term. The third term, \eqref{eq:mat_cp} and \eqref{eq:vac_cp}, is an interference term between the previous two terms and is called the CP term because of its dependence on the CP phase, $\delta$. To a good approximation, Eq.~\eqref{eq:vacuum} applies to the shortest baseline we consider, 130 km, whereas Eq.~\eqref{eq:matter} applies to the remaining baselines.
 
Considering the oscillation parameters which we wish to measure, the atmospheric term is sensitive to $\theta_{13}$ and $\theta_{23}$, the CP term is sensitive to $\theta_{13}$, $\delta$, sign($\Delta m_{31}^{2}$) (the mass hierarchy) and $\theta_{23}$, and the solar term is sensitive only to $\theta_{23}$. Since the atmospheric term has a quadratic dependence on $\theta_{13}$ whereas the CP term has a linear dependence, the atmospheric term dominates in the scenario that $\theta_{13}$ is large ($\sin^{2}2\theta_{13}\gtrsim10^{-2}$). The CP term dominates for intermediate values of $\theta_{13}$ (if $\delta$ is not close to $0$ or $\pi$). Since the dependence on $\delta$ enters via the oscillatory cosine term (which can take either a positive or negative sign, depending on the value of its phase), there can be constructive or destructive interference between the atmospheric and CP terms, which is why the sensitivities to $\theta_{13}$ and the mass hierarchy have a strong dependence on the value of $\delta$. Due to its inverse dependence on the energy, the CP term becomes most visible at \emph{lower} energies, for a fixed $L$; therefore it is important to use a detector with a low energy threshold to establish if CP is violated. The solar term, which acts as a `background' to the other two terms when performing measurements of $\theta_{13},\,\delta$ and the mass hierarchy, is dominant when $\theta_{13}$ is very small. All these statements are true for both the vacuum and matter cases.

Now we will briefly discuss the differences between the probabilities in vacuum and in matter. To begin with, the atmospheric term is modified so that it now has a dependence on sign($\Delta m_{31}^{2}$); therefore the sensitivity to the mass hierarchy is increased. For the ratio $L/E$ tuned to the first oscillation peak, this term is dominant in the probability, thus providing good sensitivities to both $\theta_{13}$ and the mass hierarchy. On the other hand, the solar term now has a dependence on $\left(\frac{\Delta_{12}}{A}\right)^{2}$ which means that it is now quadratically suppressed with the energy and/or the matter density. The CPV term is also suppressed according to the same ratio, $\left(\frac{\Delta_{12}}{A}\right)$, but the dependence in this case is linear instead of quadratic. Therefore the CP and solar terms are dominant for low energies, whereas the atmospheric term is dominant for high energies. This means that, for $L/E$ tuned to the first oscillation peak, and in the presence of strong matter effects, it should be possible to obtain excellent sensitivities to $\theta_{13}$, the mass hierarchy, and $\theta_{23}$ since the background solar term is strongly suppressed. The situation is different if we want to observe CPV though. In this case, lower energies (shorter baselines) are preferable so that the CP term is larger. In addition, the precision of the measurement is heavily affected by the presence of degeneracies; therefore it is also necessary to measure the other parameters so as to lift them. Therefore, maximising the sensitivity to CPV is a complicated combination of maximising the magnitude of the CP term by using a fairly low energy without allowing the solar term to dominate, whilst at the same time being able to determine the values of the other oscillation parameters.

\section{Event rates}
\label{sec:events}

\subsection*{Number of $\nu_\mu\rightarrow \nu_e$ events at different baselines}

In Tab.~\ref{tab:large} and Tab.~\ref{tab:largebar} we present the total number of $\nu_\mu \rightarrow \nu_e$ events obtained at the different baselines, for $\sin^2\theta_{13}=0.1$ and different values of $\delta$. The number of events correspond to one year of exposure, assuming a 100 kton detector with perfect efficiency and energy resolution. The input values for the solar and atmospheric parameters are those quoted in Sec.~\ref{sec:sim}.

{\renewcommand{\arraystretch}{1.6}
\begin{center}
 \begin{table}[htb]
  \begin{tabular}{r|c|c|c|c|c}
	       & \; 2300 km \; & \; 1570 km \; &\;  1050 km\;  & \; 665 km\;   & \; 130 km \;  \\ \hline
    $90^\circ$ & 961  & 1123 & 1386 & 1032 & 740 \\
    $0^\circ$  & 1158 & 1427 & 1780 & 1366 & 984 \\
   $-90^\circ$ & 1388 & 1659 & 1963 & 1540 & 1028 \\
  $-180^\circ$ & 1190 & 1356 & 1550 & 1205 & 784 \\
  \end{tabular}
\caption{Number of $\nu_\mu \rightarrow \nu_e $ events for the different baselines under consideration, assuming $\sin^22\theta_{13} = 10^{-1}$, and four different values of $\delta$, as indicated in the table. Normal hierarchy has also been assumed. The number of events correspond to a 100 kton detector of perfect efficiency and energy resolution, after 1 year of exposure. \label{tab:large}} 
 \end{table}
\end{center}
}

{\renewcommand{\arraystretch}{1.6}
\begin{center}
 \begin{table}[htb]
  \begin{tabular}{r|c|c|c|c|c}
	       & \; 2300 km \; & \; 1570 km \; &\;  1050 km\;  & \; 665 km\;   & \; 130 km \;  \\ \hline
    $90^\circ$ & 165  & 322 & 547 & 424 & 151 \\
    $0^\circ$  & 141 & 280 & 512 & 382 & 142 \\
   $-90^\circ$ & 83 & 185 & 367 & 274 & 106 \\
  $-180^\circ$ & 106 & 226 & 402 & 316 & 115 \\
  \end{tabular}
\caption{Number of $\bar\nu_\mu \rightarrow \bar\nu_e $ events for the different baselines under consideration, assuming $\sin^22\theta_{13} = 10^{-1}$, and four different values of $\delta$, as indicated in the table. Normal hierarchy has also been assumed. The number of events correspond to a 100 kton detector of perfect efficiency and energy resolution, after 1 year of exposure. \label{tab:largebar}} 
 \end{table}
\end{center}
}

\subsection*{Number of $\nu_\mu\rightarrow \nu_\tau$ events at different baselines}
\label{sec:apptaus}

In Tab.~\ref{tab:taus}, we present the total number of $\nu_\mu \rightarrow \nu_\tau$ events obtained at two of the baselines under consideration, for $\sin^2\theta_{13}=0.1$ and several values of $\delta$, to illustrate the discussion in Sec.~\ref{sec:taus}. The number of events correspond to one year of exposure, assuming a 100 kton detector with perfect efficiency and energy resolution. The input values for the solar and atmospheric parameters are those quoted in Sec.~\ref{sec:sim}. We show the number of events for the setup with the highest neutrino energy ($L=2300$ km) as well as for the setup with the largest statistics for the $\nu_e$ appearance channel ($L=1050$ km, see Tab.~\ref{tab:large}).

{\renewcommand{\arraystretch}{1.6}
\begin{center}
 \begin{table}[htb]
 
  \begin{tabular}{r|c|c}
   & \; 2300 km \; & \; 1050 km \; \\ \hline
    $90^\circ$  & 355 & 162 \\
    $0^\circ$   & 351 & 160  \\
   $-90^\circ$ & 347 & 159  \\
  $-180^\circ$  & 351 & 160  \\
  \end{tabular}
 
\caption{Number of $\nu_\mu \rightarrow \nu_\tau $ events for two of the baselines under consideration, assuming $\sin^22\theta_{13} = 0.1$, for four different values of $\delta$, as indicated in each row. The number of events correspond to a 100 kton detector of perfect efficiency and energy resolution, after 1 year of exposure. \label{tab:taus}} 
 \end{table}
\end{center}
}


\end{document}